\RequirePackage{rotating}
\documentclass[useAMS,usenatbib]{mnras}

\usepackage{ulem}
\usepackage{color}
\usepackage{graphics,graphicx}
\usepackage{times} 
\usepackage{amssymb}
\usepackage{amsmath}
\usepackage{natbib}
\usepackage{lscape}
\usepackage{url}
\usepackage{multirow}
\usepackage{ctable}
\usepackage{threeparttable}
\usepackage[labelfont=bf,labelsep=period]{caption}
\newif\ifAMStwofonts
\AMStwofontstrue

\def\xmm{{\it XMM-Newton}}

\def\spitzer{{\it Spitzer}}

\def\suzaku{{\it Suzaku}}

\def\swift{{\it Swift}}
\def\swiftng{{\it Neil Gehrels Swift Observatory}}
\def\integral{{\it INTEGRAL}}

\def\epicpn{{EPIC-pn}}
\def\epicmos1{{EPIC-MOS1}}
\def\epicmos2{{EPIC-MOS2}}
\def\epicmos{{EPIC-MOS}}

\def\nustar{{\it NuSTAR}}

\def\hexp{{\it HEX-P}}

\def\pcmsq{\hbox{$\rm\thinspace cm^{-2}$}}
\def\H0{{\rm ~km~s^{-1}~Mpc^{-1}}}

\def\ergpcmsqps{\hbox{$\rm\thinspace erg~cm^{-2}~s^{-1}$}}

\def\ergps{\hbox{erg~s$^{-1}$}}

\def\msun{\hbox{$M_{\odot}$}}

\def\mdot{$\dot{m}$}

\def\chisq{{$\chi^{2}$}}

\def\xspec{\hbox{\small XSPEC}}

\def\heasoft{\hbox{\rm{\small HEASOFT}}}
\def\nustardas{\rm {\small NUSTARDAS}}

\def\xmmselect{\hbox{\rm{\small XMMSELECT}}}
\def\ftool{\hbox{\rm{\small FTOOL}}}

\def\addascaspec{\hbox{\rm{\small ADDASCASPEC~\/}}}
\def\flx2xsp{\rm{\small FLX2XSP}}

\def\epchain{\hbox{\rm{\small EPCHAIN}}}
\def\emchain{\hbox{\rm{\small EMCHAIN}}}

\def\rmfgen{\hbox{\rm{\small RMFGEN}}}
\def\arfgen{\hbox{\rm{\small ARFGEN}}}

\def\addascaspec{\rm{\small ADDASCASPEC}}
\def\nupipeline{\rm{\small NUPIPELINE}}
\def\nuproducts{\rm{\small NUPRODUCTS}}

\def\simpl{\textsc{simpl}}

\def\tbabs{\textsc{tbabs}}

\def\diskbb{\textsc{diskbb}}
\def\diskpbb{\textsc{diskpbb}}

\def\cutoffpl{\textsc{cutoffpl}}

\def\eg{{\it e.g.}}

\def\ie{{\it i.e.~\/}}

\def\la{\mathrel{\hbox{\rlap{\hbox{\lower4pt\hbox{$\sim$}}}{\raise2pt\hbox{$<$}}}}}
\def\ga{\mathrel{\hbox{\rlap{\hbox{\lower4pt\hbox{$\sim$}}}{\raise2pt\hbox{$>$}}}}}

\def\d25{D$_{25}$}
\def\nh{{$N_{\rm H}$}}

\def\rsp{$R_{\rm{sp}}$}
\def\rmag{$R_{\rm{M}}$}
\def\rco{$R_{\rm{co}}$}

\def\Tin{$T_{\rm{in}}$}
\def\fsc{$f_{\rm{sc}}$}

\def\ecut{$E_{\rm{cut}}$}

\def\hoix{Holmberg\,IX X-1}
\def\hexphighTexp{60}
\def\hexplowTexp{50}
\def\hexptotexp{600}
\def\totcurrent{1.6\,Ms}

\title[HoIX X-1: Broadband Spectral Variability]{A New Broadband Spectral State in the Ultraluminous X-ray Source Holmberg IX X-1}

\author[D.\,J. Walton et al.]
{\parbox{7.in}{D.\,J. Walton$^{1}$\thanks{E-mail: d.walton4@herts.ac.uk},
M. Bachetti$^{2}$,  
P. Kosec$^{3}$,  
F. F\"urst$^{4}$,  
C. Pinto$^{5}$,  
T. P. Roberts$^{6}$,  
R. Soria$^{7,8}$,  
D. Stern$^{9}$,  
\\[0.1cm]
W. N. Alston$^{1}$, 
M. Brightman$^{10}$,  
H. P. Earnshaw$^{10}$,  
A. C. Fabian$^{11}$,  
F. A. Harrison$^{10}$,
M. J. Middleton$^{12}$,  
\\[0.1cm]
R. Sathyaprakash$^{13}$  
\\
\\[-0.2cm]
\footnotesize
$^{1}$ \it{Centre for Astrophysics Research, University of Hertfordshire, College Lane, Hatfield AL10 9AB, UK} \\
$^{2}$ \it{INAF-Osservatorio Astronomico di Cagliari, via della Scienza 5, I-09047 Selargius, Italy} \\
$^{3}$ \it{Center for Astrophysics — Harvard \& Smithsonian, Cambridge, MA, USA} \\
$^{4}$ \it{European Space Astronomy Centre (ESA/ESAC), Operations Department, Villanueva de la Ca\~nada (Madrid), Spain} \\
$^{5}$ \it{INAF - IASF Palermo, Via U. La Malfa 153, I-90146 Palermo, Italy} \\
$^{6}$ \it{Centre for Extragalactic Astronomy and Dept. of Physics, Durham University, South Road, Durham DH1 3LE, UK} \\
$^{7}$ \it{INAF-Osservatorio Astrofisico di Torino, Strada Osservatorio 20, I-10025 Pino Torinese, Italy} \\
$^{8}$ \it{Sydney Institute for Astronomy, School of Physics A28, The University of Sydney, Sydney, NSW 2006, Australia} \\
$^{9}$ \it{Jet Propulsion Laboratory, California Institute of Technology, Pasadena, CA 91109, USA} \\
$^{10}$ \it{Cahill Center for Astronomy and Astrophysics, California Institute of
Technology, 1200 E. California Boulevard, Pasadena, 91125, CA, USA} \\
$^{11}$ \it{Institute of Astronomy, University of Cambridge, Madingley Road, Cambridge CB3 0HA, UK} \\
$^{12}$ \it{Department of Physics and Astronomy, University of Southampton, Highfield, Southampton SO17 1BJ, UK} \\
$^{13}$ \it{Scuola Universitaria Superiore IUSS Pavia, Palazzo del Broletto, piazza della Vittoria 15, I-27100 Pavia, Italy}
}}
\date{}

\begin{document}
\pagerange{\pageref{firstpage}--\pageref{lastpage}}
\maketitle
\label{firstpage}

\begin{abstract}
We present a series of five new broadband X-ray observations of the ultraluminous
X-ray source \hoix, performed by \xmm\ and \nustar\ in coordination.  The first three
of these show high soft X-ray fluxes but a near total collapse of the high-energy
($\gtrsim$15\,keV) emission, previously seen to be surprisingly stable across all prior
broadband observations of the source. The latter two show a recovery in hard X-rays,
remarkably once again respecting the same stable high-energy flux exhibited by all
of the archival observations. We also present a joint analysis of all broadband
observations of \hoix\ to date (encompassing 11 epochs in total) in order to
investigate whether it shows the same luminosity--temperature behaviour as
NGC\,1313 X-1 (which also shows a stable high-energy flux), whereby the hotter disc
component in the spectrum exhibits two distinct, positively-correlated tracks in the
luminosity--temperature plane. \hoix\ may show similar behaviour, but the results
depend on whether the highest energy emission is assumed to be an up-scattering
corona or an accretion column. The strongest evidence for this behaviour is found in
the former case, while in the latter the new `soft' epochs appear distinct from the
other high-flux epochs. We discuss possible explanations for these new `soft'
spectra in the context of the expected structure of super-Eddington accretion flows
around black holes and neutron stars, and highlight a potentially interesting analogy
with the recent destruction and re-creation of the corona seen in the AGN
1ES\,1927+654.
\end{abstract}

\begin{keywords}
{X-rays: Binaries -- X-rays: individual (\hoix)}
\end{keywords}

\section{Introduction}

The \nustar\ era has brought about significant progress in our understanding of the
ultraluminous X-ray source (ULX) population. Seen almost exclusively in other galaxies,
these are the most extreme members of the X-ray binary population and exhibit
luminosities in excess of $10^{39}$\,\ergps\ (the Eddington limit for a typical
$\sim$10\,\msun\ stellar remnant black hole); see \cite{King23rev} and \cite{Pinto23rev}
for recent reviews. Explanations for these extreme luminosities in the literature typically
focused on either sub-Eddington accretion onto intermediate mass black holes (IMBHs,
$M_{\rm{BH}} \sim 10^{2-5}$\,\msun; \eg\ \citealt{Miller03, Strohmayer09a}) or
super-Eddington accretion onto relatively normal stellar remnants (\eg\ \citealt{King01,
Poutanen07}). However, we now understand the majority of ULXs to represent the best
local examples of the latter. \nustar\ (\citealt{NUSTAR}) has enabled detailed studies of
ULXs in the hard X-ray band ($E \gtrsim 10$\,keV) for the first time, revealing
broadband spectra inconsistent with standard modes of sub-Eddington accretion (\eg\
\citealt{Bachetti13, Walton13culx, Walton15, Walton15hoII, Rana15, Mukherjee15,
Earnshaw19, Brightman22}), confirming prior indications seen at lower energies (\eg\
\citealt{Stobbart06, Gladstone09}). More importantly, \nustar\ observations of the M82
galaxy resulted in the discovery of the first ULX pulsar (\citealt{Bachetti14nat}),
unambiguously identifying the accretor as a highly super-Eddington neutron star. Since
this discovery, at least five more ULX pulsars have been revealed (\citealt{Fuerst16p13,
Israel17p13, Israel17, Carpano18, Sathyaprakash19, Rodriguez20, Pintore25}).
High-resolution spectroscopy with \xmm\ (\citealt{XMM, XMM_RGS}) has also recently
revealed the presence of powerful outflows in a number of ULXs (\citealt{Pinto16nat,
Pinto17, Pinto20, Walton16ufo, Kosec18ulx, Kosec18, Kosec21}), further cementing the
association between ULXs and super-Eddington accretion as such winds are a
ubiquitous prediction for this regime.

\hoix\ is one of the better studied members of the ULX population, although as no
X-ray pulsations have ever been seen from this source (\citealt{Doroshenko15,
Walton17hoIX}; Appendix \ref{app_pulse}) the nature of the accretor remains unknown.
Located at a distance of 3.55\,Mpc (\citealt{Paturel02}), it is one of the nearest ULXs
that almost persistently exhibits luminosities in excess of $10^{40}$\,\ergps. Studies of
\hoix\ have generally mirrored the same evolution as studies of the broader ULX
population. Early observations with \xmm\ revealed potential evidence for a
low-temperature accretion disc ($kT \sim 0.2-0.3$\,keV, consistent with the idea that
this may host a $\sim$1000\,\msun\ IMBH (\citealt{Miller03}). However, higher
signal-to-noise (S/N) data subsequently showed evidence that the emission in the
2--10\,keV band was not a standard sub-Eddington powerlaw continuum (\eg\
\citealt{Gladstone09, Walton13hoIXfeK}), which was further confirmed by high-energy
observations with \integral\ (\citealt{Sazonov14}) and with coordinated broadband
observations combining \xmm, \suzaku\ (\citealt{SUZAKU}) and \nustar\
(\citealt{Walton14hoIX}; see also \citealt{Luangtip16}). Further broadband follow-up
combining \suzaku\ and \nustar\ showed unusual spectral variability: strong variations
were seen between observing epochs at lower energies ($\lesssim$10\,keV),
with the peak of the X-ray spectrum shifting to lower energies at higher fluxes,
while the high-energy flux ($\gtrsim$15\,keV) remained comparatively stable
(\citealt{Walton17hoIX}; see also \citealt{Gurpide21}). Similar behaviour has also been
seen in multi-epoch broadband observations of another well-studied ULX,
NGC\,1313 X-1 (\citealt{Walton20}), but its origin is not currently well understood. 

With the availability of broadband data for ULXs in the \nustar\ era, standard
spectral models for these sources have come to include two thermal accretion disc
components (typical temperatures $\sim$0.2--0.5 and $\sim$1--3\,keV; \eg\ 
\citealt{Walton14hoIX, Walton18ulxBB, Koliopanos17}) -- the cooler of which may be
associated with a combination of emission from the outer disc and the large wind
expected for super-Eddington accretion, while the hotter may be associated with the
innermost, funnel-like regions of a super-Eddington disc -- and an additional steep
high-energy continuum component (either from an accretion column or an
up-scattering corona; \eg\ \citealt{Walton15, Walton18ulxBB, Mukherjee15}).  When
the multi-epoch data for NGC\,1313 X-1 were fit with these models (the nature of the
accretor in NGC\,1313 X-1 is also not yet known) the evolution of the hotter disc
component exhibited some unusual behaviour in the luminosity--temperature plane.
The broadband observations showed two distinct groups, consisting of a set with
lower fluxes and higher temperatures, and a set with higher fluxes but lower
temperatures. However, both of these individual groups showed evidence for having
its own \textit{positive} luminosity--temperature relation (i.e. higher temperatures at
higher fluxes), such that the inner disc component appeared to be following two
distinct luminosity--temperature tracks. This strange luminosity--temperature
behaviour is itself challenging to understand, but also made it even more difficult to
understand the relative consistency of the highest energy data.

While the nature of this spectral evolution is still not clear, it is important to establish
whether the same behaviour is seen in other sources. As noted above, the same kind
of evolution appears to be present in \hoix, but the archival data only include one
observation that would correspond to the high-flux, low-temperature group seen in
NGC\,1313 X-1. 

Here we present a set of five new broadband observations of \hoix\ with \xmm\ and
\nustar\ that allow us to further explore the strange spectral variability seen from
this source. The main result from these new observations is the revelation of a new,
soft broadband spectral `state' -- seen during the first three of these new
observations -- in which the previously stable high-energy flux has collapsed
(though the source does subsequently recover to the same high-energy flux in the
latter two). The rest of the paper is structured as follows: the new observations and
our data reduction are described in Section \ref{sec_red}, and our spectral analysis
is presented in Section \ref{sec_spec}. The results from this analysis are discussed
in Section \ref{sec_dis} (including a detailed consideration of the potential nature of
this new soft state in Section \ref{sec_soft_dis}),  a brief exploration into the
advances that a mission similar to the recently-proposed \textit{High Energy X-ray
Probe} (\textit{HEX-P}; \citealt{HEXP_2024}) would provide for broadband studies
of ULXs is presented in section \ref{sec_hexp}, and finally our main conclusions are
summarised in Section \ref{sec_conc}.

\section{Observations and Data Reduction}
\label{sec_red}

During the course of late 2020 (October -- November), \xmm\ and \nustar\ performed
a series of five new coordinated observations of \hoix.\footnote{Four new observations
were originally planned, but during the third \xmm\ exposure (OBSID 0870930401) the
\epicmos\ detectors experienced a fault, and so a fifth observation was scheduled.
However, the \epicpn\ detector was still fully operational during this third observation,
and the \epicmos\ detectors did operate for the last $\sim$25--30\% of the exposure,
so we still analyse the data that is available for this observation.} In total, there
are now 11 epochs of coordinated broadband X-ray observations of \hoix\ taken with
some combination of \nustar, \xmm\ and \suzaku. These new observations (broadband
epochs 7--11) were triggered on the transition to a high-flux state below 10\,keV,
detected by monitoring with the \swiftng\ (hereafter \swift; \citealt{SWIFT}); we show
them in the context of the long-term evolution of \hoix\ seen by the \swift\ XRT
(\citealt{SWIFT_XRT}) in Figure \ref{fig_longlc}. The following sections describe these
new observations in more detail, and outline our data reduction procedure. Basic
information for all the broadband observations considered in this work is given in
Table \ref{tab_obs}.

\begin{table}
  \caption{Details of the broadband observations of \hoix\ used in this work (given in
  chronological order).}
\begin{center}
\begin{tabular}{c c c c c c}
\hline
\hline
\\[-0.25cm]
Epoch & Mission(s) & OBSID(s) & Start & Exposure\tmark[a] \\
\\[-0.3cm]
& & & Date & (ks) \\
\\[-0.25cm]
\hline
\hline
\\[-0.2cm]
\multirow{8}{*}{1} & \suzaku\ & 707019020 & 2012-10-21 & 107 \\
\\[-0.3cm]
& \xmm\ & 0693850801 & 2012-10-23 & 10/14 \\
\\[-0.3cm]
&  \suzaku\ & 707019030 & 2012-10-24 & 107 \\
\\[-0.3cm]
& \xmm\ & 0693850901 & 2012-10-25 & 11/14 \\
\\[-0.3cm]
& \suzaku\ & 707019040 & 2012-10-26 & 110 \\
\\[-0.3cm]
& \nustar\ & 30002033002 & 2012-10-26 & 43 \\
\\[-0.3cm]
& \nustar\ & 30002033003 & 2012-10-26 & 124 \\
\\[-0.3cm]
& \xmm\ & 0693851001 & 2012-10-27 & 11/13 \\
\\
\multirow{7}{*}{2} & \nustar\ & 30002033005 & 2012-11-11 & 49 \\
\\[-0.3cm]
& \nustar\ & 30002033006 & 2012-11-11 & 41 \\
\\[-0.3cm]
& \xmm\ & 0693851701 & 2012-11-12 & 7/10 \\
\\[-0.3cm]
& \nustar\ & 30002033008 & 2012-11-14 & 18 \\
\\[-0.3cm]
& \xmm\ & 0693851801 & 2012-11-14 & 10//9 \\
\\[-0.3cm]
& \nustar\ & 30002033010 & 2012-11-15 & 59 \\
\\[-0.3cm]
& \xmm\ & 0693851101 & 2012-11-16 & 10/13 \\
\\
\multirow{2}{*}{3} & \nustar\ & 30002034002 & 2014-05-02 & 81 \\
\\[-0.3cm]
& \suzaku\ & 707019010 & 2014-05-03 & 32 \\
\\
\multirow{2}{*}{4} & \nustar\ & 30002034004 & 2014-11-15 & 81 \\
\\[-0.3cm]
& \suzaku\ & 707019020 & 2014-11-15 & 34 \\
\\
\multirow{2}{*}{5} & \nustar\ & 30002034006 & 2015-04-06 & 64 \\
\\[-0.3cm]
& \suzaku\ & 707019030 & 2015-04-06 & 32 \\
\\
\multirow{2}{*}{6} & \nustar\ & 30002034008 & 2015-05-16 & 67 \\
\\[-0.3cm]
& \suzaku\ & 707019040 & 2015-05-16 & 34 \\
\\
\multirow{2}{*}{7} & \nustar\ & 80602308002 & 2020-10-17 & 66 \\
\\[-0.3cm]
& \xmm\ & 0870930101 & 2020-10-17 & 16/22 \\
\\
\multirow{2}{*}{8} & \nustar\ & 80602308004 & 2020-10-29 & 66 \\
\\[-0.3cm]
& \xmm\ & 0870930301 & 2020-10-29 & 16/14 \\
\\
\multirow{2}{*}{9} & \nustar\ & 80602308006 & 2020-11-05 & 76 \\
\\[-0.3cm]
& \xmm\ & 0870930401 & 2020-11-06 & 17/7\tmark[b] \\
\\
\multirow{2}{*}{10} & \nustar\ & 80602308008 & 2020-11-18 & 64 \\
\\[-0.3cm]
& \xmm\ & 0870930501 & 2020-11-18 & 18/25 \\
\\
\multirow{2}{*}{11} & \nustar\ & 80602308010 & 2020-11-23 & 69 \\
\\[-0.3cm]
& \xmm\ & 0870931001 & 2020-11-24 & 17/24 \\
\\[-0.2cm]
\hline
\hline
\\[-0.4cm]
\end{tabular}
\end{center}
$^{a}$ \xmm\ exposures are listed separately for the \epicpn/MOS detectors after
background filtering, and the \nustar\ exposures combine both modes 1 and 6; all
exposures are given to the nearest ks. \\
$^{b}$ The \epicmos\ detectors experienced a fault during this observation, hence the
reduced exposure.
\label{tab_obs}
\end{table}

\subsection{\textit{NuSTAR}}

We reduced the \nustar\ data following standard procedures with the \nustar\ Data
Analysis Software (v2.0.0; part of the \heasoft\ distribution) and \nustar\ caldb
v20211202. The raw event files for each observation were initially cleaned with
\nupipeline, using the standard depth correction to reduce the internal background.
Passages through the South Atlantic Anomaly and periods during which \hoix\ was
occulted by the earth were also excluded. Source products were extracted from the
cleaned event files for each focal plane module (FPMA, FPMB) using \nuproducts. We
used circular regions of radius $\sim$75$''$ for the source aperture, and the
background was estimated from a larger, blank area on the same detector, free of
contaminating point sources. In addition to the standard `science' data (mode 1), we
also extract the `spacecraft science' data (mode 6) in order to maximise the
signal-to-noise (S/N), following \cite{Walton16cyg}. This provides $\sim$20--40\% of
the total good exposure, depending on the specific observation.

\subsection{\textit{XMM-Newton}}

The \xmm\ observations were also reduced following standard procedures with the
\xmm\ Science Analysis System (v19.1.0). For the \epicpn\ and \epicmos\ CCD
detectors (\citealt{XMM_PN, XMM_MOS}) we processed the raw data files for each
observation using \epchain\ and \emchain\ to produce calibrated event lists. For
these new \xmm\ exposures, each of the EPIC detectors was operated in Small Window
mode. Source products were extracted from the cleaned event files using \xmmselect,
using circular regions of radius $\sim$35$''$ for the source aperture. For the \epicpn\
detector the background was always estimated from a larger region of blank sky on the
same CCD as the source. This was not possible for the \epicmos\ detectors owing to the
use of Small Window mode, so for these detectors background was estimated from
larger regions of blank sky on adjacent chips. As recommended, we only consider single
and double patterned events for the \epicpn\ detector ({\small PATTERN}$\leq$4) and
single to quadruple patterned events for the \epicmos\ detectors
({\small PATTERN}$\leq$12), and for \epicpn\ we only consider events with {{\small
FLAG}} = 0. Periods of high background were excluded from our analysis, and we
determined the appropriate background threshold to exclude based on the technique
outlined by \cite{Picon04}, which determines the background level that maximises the
S/N for the source data (we use the full 0.3--10.0\,keV EPIC bandpass for this
assessment). The instrumental response files for each of the EPIC detectors were
generated with \rmfgen\ and \arfgen. After performing the data reduction separately
for each of the \epicmos\ units, and confirming their consistency, the spectra from
these detectors were combined using the \ftool\ \addascaspec\ for each observation.

\subsection{Archival Broadband Observations}

In addition to these new observations, we also consider the six archival broadband
observations combining \xmm, \suzaku\ and \nustar\ reported by \cite{Walton14hoIX}
and \cite{Walton17hoIX}. The data reduction for these observations largely follows the
process outlined in those works, but where relevant we have reprocessed the data with
updated instrumental calibration files (the \suzaku\ calibration has not been updated
since the latter of those works were published),  and have updated the reduction of
the archival \xmm\ observations to also utilize the \cite{Picon04} technique
highlighted above to maximise the S/N of the source data in the presence of any
background flaring.

\begin{figure*}
\begin{center}
\hspace*{-0.3cm}
\rotatebox{0}{
{\includegraphics[width=500pt]{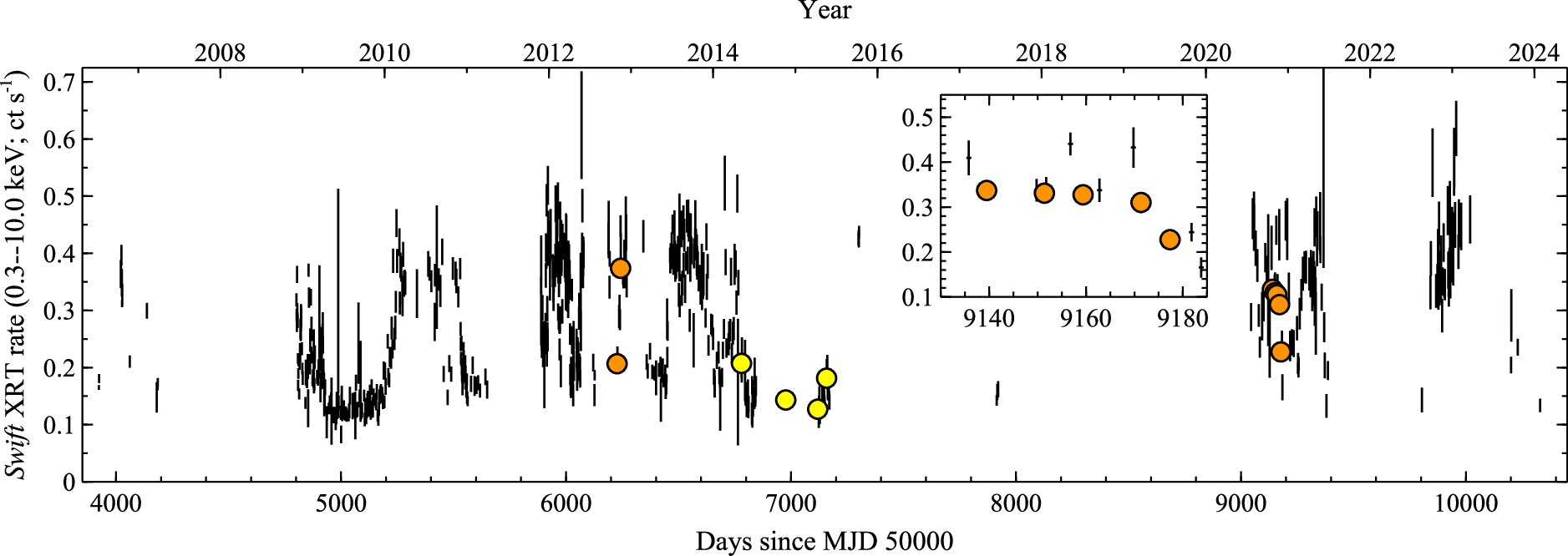}}
}
\end{center}
\vspace*{-0.3cm}
\caption{
The long-term X-ray lightcurve of \hoix\ seen by \swift/XRT (1d time bins), extracted
using the online pipeline (\citealt{Evans09}). The broadband observations considered
here are shown with the large orange/yellow circles, after converting the
\xmm/\suzaku\ data to equivalent XRT count rates based on their observed spectra.
The inset shows a zoom-in on the most recent series of broadband observations.}
\label{fig_longlc}
\end{figure*}

\section{Analysis}
\label{sec_spec}

With these new observations in hand, we naturally undertook an updated timing analysis
to search for any coherent pulsations that would identify \hoix\ as another member of
the ULX pulsar family. However, we did not detect any compelling signals (see Appendix
\ref{app_pulse}), so we still consider the nature of the accretor in \hoix\ to be uncertain.
We therefore focus our main analysis on the broadband spectral properties of the new
observations, and the spectral variability exhibited when compared to the archival
observations. Our spectral analysis is performed with \xspec\ (version 12.11.1;
\citealt{xspec}) and we fit the data by reducing the \chisq\ 
statistic; all the datasets considered here are binned to have a minimum S/N of 5 to
facilitate this. All our spectral models include absorption from the column through our
own Galaxy in the direction of \hoix\ ($N_{\rm{H,Gal}} = 5.5 \times 10^{20}$\,\pcmsq;
\citealt{NH2016}), in addition to neutral absorption more local to the source. Both are
modelled with the \tbabs\ absorption code (\citealt{tbabs}), and we adopt the solar
abundances quoted in that work as well as the absorption cross-sections of
\cite{Verner96}, as recommended for this model. Finally, unless stated otherwise,
parameter uncertainties are given at the 90\% level (i.e. $\Delta\chi^{2} = 2.71$).

Of the five new observations (epochs 7--11), the last two show relatively similar
spectra to prior broadband observations of \hoix; epoch 10 is reasonably similar to
the `high' state observation discussed in \cite{Walton17hoIX} (epoch 2), and epoch
11 is similar to the `medium' state observations discussed in that same work (epochs
1, 3 and 6). However, as noted above, epochs 7--9 are distinctly different from prior
broadband observations (although are themselves all relatively similar to one
another). The high-energy flux -- which previously appeared to be remarkably stable
across epochs 1--6 despite the strong variations seen below $\sim$10\,keV -- has
significantly collapsed in these three observations. We show a comparison of the
broadband spectrum seen from the first of these (epoch 7) with the three prior
representative broadband spectra highlighted by \cite{Walton17hoIX} in Figure
\ref{fig_spec_4states}. At low energies (below $\sim$2\,keV) these new spectra
initially appear to be similar to the high state data, but they then peak at lower energy
($\sim$3--4\,keV) before falling away very sharply; the source is generally not
detected above $\sim$20\,keV in these observations (in contrast to earlier
observations, in which the source was detected up to $\sim$40\,keV).

\subsection{A New Broadband Spectral State}
\label{sec_soft}

In order to characterise this new `soft' broadband spectral state\footnote{Note that we
are using this nomenclature purely to distinguish these new observations of \hoix\ from
the broadband spectra reported previously (\citealt{Walton14hoIX, Walton17hoIX}); we
are not implying that during these observations \hoix\ is in the classical soft state seen
in sub-Eddington X-ray binaries (\eg\ \citealt{Remillard06rev}).} we focus initially on
the data from epoch 7, as the first of these three relevant observations and also the one
with the highest S/N \nustar\ data. Given the significant collapse in the high-energy flux
seen during this soft state, we initially consider a model that consists of just the two
thermal components highlighted previously. Specifically, we start by combining the
\diskbb\ and \diskpbb\ models (\citealt{diskbb, diskpbb}), as has become fairly
standard within the recent literature. \diskbb\ formally describes the thermal emission
from a standard geometrically thin accretion disc (which should be present outside of
the `spherisation' radius, \rsp, the point at which the total luminosity integrated from 
the outer disc inwards reaches the Eddington luminosity; \citealt{Shakura73}), but its
thermal nature likely means it is also well suited to describing any cooler emission from
an optically-thick wind; this model is simply characterised by an inner disc temperature
(\Tin) and a normalisation. The \diskpbb\ model is slightly more complex as it also
allows the radial temperature index ($p$) to vary as an additional free parameter, and
so is often used as a simple approximation of a thicker super-Eddington accretion disc
(which would be expected interior to \rsp, provided that the disc has not been
truncated outside of this point should the accretor be a magnetic neutron star).

\begin{figure}
\begin{center}
\hspace*{-0.2cm}
\rotatebox{0}{
{\includegraphics[width=240pt]{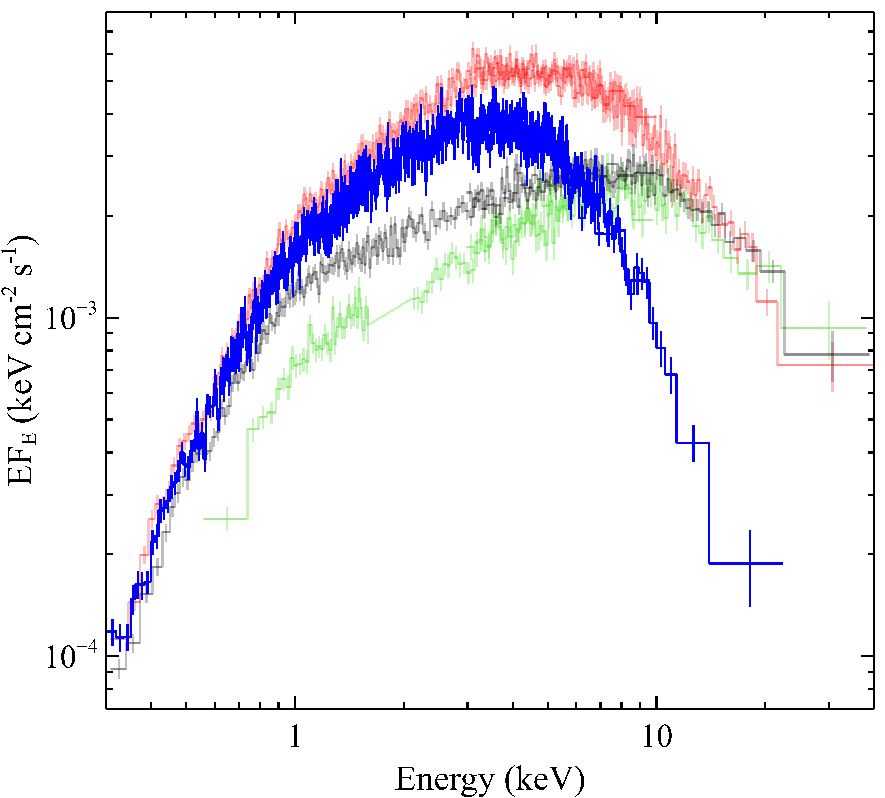}}
}
\end{center}
\vspace*{-0.3cm}
\caption{
The broadband spectrum of one of the new soft state observations of \hoix\ (epoch 7;
solid blue), compared to the three broadband spectra highlighted by \citet{Walton17hoIX}
that are representative of the archival observations (epochs 1, 2 and 4; faded black, red
and green, respectively). For clarity, we only show the \nustar\ data from FPMA along
with one of the accompanying soft X-ray detectors (either the \epicpn\ data for \xmm\ or
the FI XIS data for \suzaku). Of the four other new broadband epochs presented in this
work, epochs 8 and 9 are similar to epoch 7 (blue), epoch 10 is similar to epoch 2 (red),
and epoch 11 is similar to epoch 1 (black). All the data have been unfolded through the
same model, which is simply constant with energy, and the data have been further
rebinned for visual purposes.}
\label{fig_spec_4states}
\end{figure}

\begin{figure}
\begin{center}
\hspace*{-0.2cm}
\rotatebox{0}{
{\includegraphics[width=240pt]{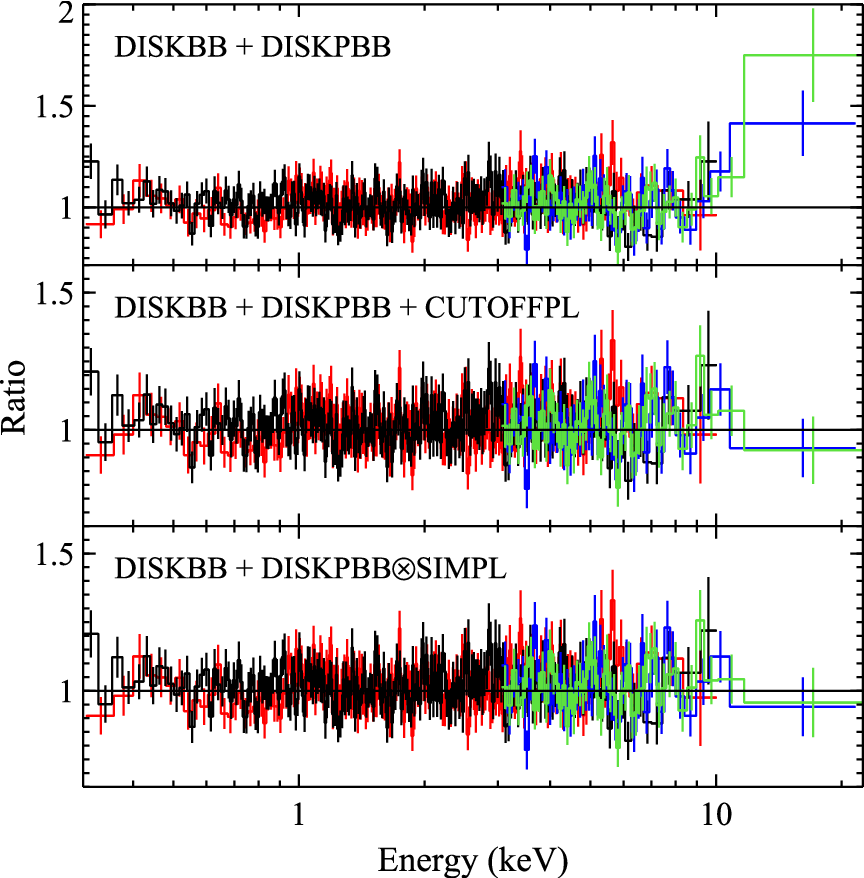}}
}
\end{center}
\vspace*{-0.3cm}
\caption{
Data/model ratios for the continuum models applied to the broadband data from epoch
7 (one of the observations showing the new `soft' state revealed by our recent
broadband campaign) presented in Table \ref{tab_soft_param}. Even though the
high-energy flux has largely collapsed here, models that only include thermal emission
(top) still leave statistically significant residuals at the highest energies seen by \nustar,
suggesting there is still some small contribution from a higher energy component
during this epoch (modelled as an accretion column and an up-scattering corona in the
middle and bottom panels, respectively). Here, the \epicpn, \epicmos, FPMA and FPMB
data are shown in black, red, green and blue, respectively, and the data have again been
further rebinned for visual purposes.
}
\label{fig_soft_ratio}
\end{figure}

This model fits the broadband data quite well in a statistical sense, with \chisq\ = 1995
for 1870 degrees of freedom (DoF), but despite the overall collapse of the high-energy
flux there are still excess residuals seen at the highest energies probed by \nustar\
during this epoch ($\sim$20\,keV, see Fig. \ref{fig_spec_4states}). We present the
best-fit spectral parameters in Table \ref{tab_soft_param} and show the data/model
ratio in Figure \ref{fig_soft_ratio}. The high-energy excess seen in the ratio plot is
qualitatively similar to what is seen in other ULXs with high quality broadband
coverage when fit with thermal models (including previous observations of \hoix), and
suggests that even during this epoch the high-energy spectrum does not fall away
with a thermal/Wien spectrum.

We therefore also explore a couple of additional models where a third, high-energy
continuum component is included as well, intended to represent both the Compton
up-scattering corona and accretion column possibilities discussed above (as the
nature of the accretor in \hoix\ is still not known). For the former scenario, we use the
\simpl\ convolution model (\citealt{simpl}), which adds on a high-energy powerlaw
continuum to a given input emission component, and mimics a scattering process by
assuming the photon number is conserved when this additional continuum is
generated. This is applied to the \diskpbb\ component, and is characterised with a
photon index ($\Gamma$) and a scattered fraction (\fsc, which essentially acts as a
normalisation for the powerlaw tail). For the latter scenario, following \cite{Walton20}
we approximate the accretion column with a \cutoffpl\ component (\ie\ a powerlaw with
an exponential high-energy cutoff) with parameters $\Gamma = 0.59$ and \ecut\ =
7.9\,keV; this is based on the average spectral form of the accretion column seen in the
other ULX pulsars for which phase-resolved spectroscopy has been possible
(\citealt{Brightman16m82x2, Walton18p13, Walton18ulxBB, Walton18crsf}); only the
normalisation of this component is free to vary.

Both models provide similarly good fits to the data; the corona model gives \chisq/DoF
= 1957/1868 and the accretion column model gives \chisq/DoF = 1962/1869.
Furthermore, in both cases the improvements in the fit are sufficiently large to indicate
that a high-energy component is still significantly detected during this epoch; the
corona model gives an improvement of $\Delta\chi^{2} = 38$ for two extra free
parameters, and the accretion column model gives an improvement of $\Delta\chi^{2}
= 34$ for one extra free parameter. In the former case the photon index is very steep
($\Gamma > 2.8$), similar to the results found when a similar model is applied to other
ULXs (\eg\ \citealt{Walton15, Mukherjee15}), and also consistent with the high-energy
photon indices found in the archival \nustar\ observations of \hoix\
(\citealt{Walton14hoIX, Walton17hoIX, Luangtip16, Gurpide21}). The scattered fraction,
however, is significantly lower for the epoch 7 data than for the archival \nustar\
observations. In the accretion column model, the flux of this component is also
significantly lower than the archival \nustar\ observations; for comparison, the
2--10\,keV flux of the \cutoffpl\ component here is $3.6^{+0.8}_{-1.1} \times
10^{-13}$\,\ergpcmsqps\ (compared with $\sim$2 $\times 10^{-12}$\,\ergpcmsqps\
when this model is applied to the archival data; \citealt{Walton18ulxBB}).

\begin{table}
  \caption{Key parameters obtained for the various continuum model fits to the high-flux data available for \hoix}
\begin{center}
\begin{tabular}{c c c c c c}
\hline
\hline
\\[-0.2cm]
Model & \multicolumn{2}{c}{Parameter} & Epoch 7 & Stacked\\
\\[-0.35cm]
Component & & & & soft state \\
\\[-0.25cm]
\hline
\hline
\\[-0.15cm]
\multicolumn{5}{c}{\textit{Continuum Model:} \diskbb+\diskpbb} \\
\\[-0.15cm]
\tbabs\ & $N_{\rm{H}}$ & [$10^{21}~\rm{cm}^{-2}$] & $1.1^{+0.2}_{-0.1}$ & $1.2 \pm 0.1$ \\
\\[-0.25cm]
\diskbb\ & $T_{\rm{in}}$ & [keV] & $0.32 \pm 0.06$ & $0.31^{+0.04}_{-0.03}$ \\
\\[-0.25cm]
& Norm &  & $7.0^{+8.0}_{-3.2}$ & $9.9^{+4.9}_{-3.0}$ \\
\\[-0.25cm]
\diskpbb\ & $T_{\rm{in}}$ & [keV] & $1.56^{+0.03}_{-0.04}$ & $1.44 \pm 0.02$ \\
\\[-0.25cm]
& $p$ & & $0.69^{+0.05}_{-0.03}$ & $0.71 \pm 0.03$ \\
\\[-0.25cm]
& Norm &  & $0.09 \pm 0.02$ & $0.13 \pm 0.02 $ \\
\\[-0.25cm]
$\chi^{2}$/DoF & & & 1995/1870 & 2529/2372 \\
\\[-0.2cm]
\hline
\hline
\\[-0.15cm]
\multicolumn{5}{c}{\textit{Continuum Model:} \diskbb+\diskpbb$\otimes$\simpl} \\
\\[-0.15cm]
\tbabs\ & $N_{\rm{H}}$ & [$10^{21}~\rm{cm}^{-2}$] & $1.0 \pm 0.1$ & $1.1 \pm 0.1$ \\
\\[-0.25cm]
\diskbb\ & $T_{\rm{in}}$ & [keV] & $0.39^{+0.02}_{-0.05}$ & $0.34^{+0.03}_{-0.02}$ \\
\\[-0.25cm]
& Norm &  & $7.5^{+3.3}_{-1.9}$ & $9.8^{+2.4}_{-1.0}$ \\
\\[-0.25cm]
\diskpbb\ & $T_{\rm{in}}$ & [keV] & $1.31^{+0.09}_{-0.04}$ & $1.34^{+0.05}_{-0.01}$ \\
\\[-0.25cm]
& $p$ & & $>0.81$ & $0.80^{+0.16}_{-0.06}$ \\
\\[-0.25cm]
& Norm &  & $0.38^{+0.06}_{-0.16}$ & $0.22^{+0.01}_{-0.06}$ \\
\\[-0.25cm]
\simpl & $\Gamma$ & & $>2.8$ & $>3.0$ \\
\\[-0.25cm]
& \fsc\ & [\%] & $15^{+4}_{-9}$ & $6^{+2}_{-5}$ \\
\\[-0.25cm]
$\chi^{2}$/DoF & & & 1957/1868 & 2502/2370 \\
\\[-0.2cm]
\hline
\hline
\\[-0.15cm]
\multicolumn{5}{c}{\textit{Continuum Model:} \diskbb+\diskpbb+\cutoffpl} \\
\\[-0.15cm]
\tbabs\ & $N_{\rm{H}}$ & [$10^{21}~\rm{cm}^{-2}$] & $1.0^{+0.2}_{-0.1}$ & $1.1 \pm 0.1$ \\
\\[-0.25cm]
\diskbb\ & $T_{\rm{in}}$ & [keV] & $0.38^{+0.05}_{-0.06}$ & $0.34 \pm 0.03$ \\
\\[-0.25cm]
& Norm &  & $7.1^{+3.9}_{-2.0}$ & $9.5^{+3.5}_{-2.3}$ \\
\\[-0.25cm]
\diskpbb\ & $T_{\rm{in}}$ & [keV] & $1.39^{+0.07}_{-0.04}$ & $1.37 \pm 0.03$ \\
\\[-0.25cm]
& $p$ & & $>0.75$ & $0.78^{+0.09}_{-0.05}$ \\
\\[-0.25cm]
& Norm &  & $0.25^{+0.06}_{-0.08}$ & $0.20^{+0.06}_{-0.05}$ \\
\\[-0.25cm]
\cutoffpl\ & $\Gamma$ & & 0.59* & 0.59* \\
\\[-0.25cm]
& \ecut\ & [keV] & 7.9* & 7.9* \\
\\[-0.25cm]
& Norm & [$10^{-5}$] & $3.5^{+0.7}_{-1.0}$ & $1.4^{+0.4}_{-0.5}$ \\
\\[-0.25cm]
$\chi^{2}$/DoF & & & 1962/1869 & 2504/2371 \\
\\[-0.2cm]
\hline
\hline
\\[-0.15cm]
\end{tabular}
\label{tab_soft_param}
\end{center}
\vspace{-0.25cm}
* Parameter was fixed during the fit; these are based on the average parameters
found for the accretion columns in the known ULX pulsars via phase-resolved
spectroscopy (see text). \\
\end{table}

One oddity with the results for this new soft state when compared with the archival
broadband data is that the radial temperature index is fairly steep in both of our
three-component models ($p > 0.75$). For comparison, a standard thin disc is should
have $p = 0.75$, and a super-Eddington disc is broadly expected to have $p < 0.75$,
as found in the similar fits to the archival data. This implies that the model is trying to
make the \diskpbb\ component much more peaked than in the archival fits. Given this,
we explore whether two thermal components are really needed in this soft state, or
whether this could be a result of the fit trying to model a single broad thermal
component with two different components, and remove the cooler \diskbb\ component
from both of the corona and the accretion column models. In both cases, doing so
degrades the fit by $\Delta\chi^{2} \sim 45$ for two fewer free parameters, suggesting
that a dual-thermal continuum below 10\,keV is still statistically preferred. It is also
notable that the absorption column local to \hoix\ inferred here is lower than the
average column found across the archival observations of \hoix\ where similar models
have been applied (\nh\ $\sim$ 1.5 $\times 10^{21}$\,\pcmsq; \citealt{Miller13ulx,
Walton17hoIX}). This could impact the radial temperature index inferred for the
\diskpbb\ component as this parameter primarily impacts the slope of its spectrum
below the peak temperature. Indeed, if we force this column density to be 1.5 $\times
10^{21}$~cm$^{-2}$ we do find the best fit values shifts to $p \sim 0.7$ in both
models (although the quality of the fit does degrade slightly).

We have also repeated this analysis after stacking all the soft state data together
(\ie co-adding epochs 7--9 using \addascaspec). These results are also presented in
Table \ref{tab_soft_param}, and are broadly similar to the results found for epoch 7
by itself. The improvement in the fit after including an additional high-energy
component still suggests this component is present in the data, but remains broadly
at the same level as when considering epoch 7 alone ($\Delta\chi^{2} \sim 25-30$).
This is because the high-energy flux in epochs 8 and 9 is even weaker than in epoch
7, and so the preference for this third component is driven almost exclusively by the
data from this epoch. In contrast, the preference for two thermal components becomes
even stronger in the stacked data; removing the \diskbb\ component results in an even
larger degradation in the fit ($\Delta\chi^{2} > 100$).

\begin{sidewaystable*}
  \caption[labelfont=bf]{Best-fit parameters from the eleven broadband spectra currently
  available for Holmberg\,IX X-1 for the models assuming a non-magnetic and a magnetic
  accretor, respectively}
  \vspace{-0.1cm}
\begin{center}
\hspace*{-0.4cm}
\resizebox{0.98\linewidth}{!}{\begin{tabular}{c c c c c c c c c c c c c c}
\hline
\hline
\\[-0.2cm]
Model & \multicolumn{2}{c}{Parameter} & & & & & & Broadband Epoch \\
\\[-0.3cm]
Component & & & 1 & 2 & 3 & 4 & 5 & 6 & 7 & 8 & 9 & 10 & 11 \\
\\[-0.3cm]
\hline
\hline
\\[-0.2cm]
\multicolumn{14}{c}{Non-Magnetic Accretor Model: \tbabs $\times$ $($ \diskbb\ $+$ \diskpbb\ $\otimes$ \simpl\ $)$} \\
\\[-0.25cm]
\tbabs\ & $N_{\rm{H,int}}$\tmark[a] & [$10^{21}$ cm$^{-2}$] & $1.34 \pm 0.05$ & -- & -- & -- & -- & -- & -- & -- & -- & -- & -- \\
\\[-0.3cm]
\diskbb & $T_{\rm{in}}$ & [keV] & $0.31 \pm 0.01$ & $0.27 \pm 0.03$ & $0.27^{+0.02}_{-0.03}$ & $0.31^{+0.04}_{-0.02}$ & $0.31^{+0.03}_{-0.02}$ & $0.45^{+0.07}_{-0.08}$ & $0.29 \pm 0.03$ & $0.31 \pm 0.03$ & $0.28 \pm 0.03$ & $0.33 \pm 0.02$ & $0.33 \pm 0.02$ \\
\\[-0.3cm]
& Norm & & $9.6^{+2.0}_{-1.6}$ & $14.3^{+6.9}_{-5.5}$ & $26.1^{+12.2}_{-8.3}$ & $9.0^{+3.8}_{-2.6}$ & $9.3^{+3.8}_{-2.6}$ & $1.0^{+0.9}_{-0.7}$ & $18.4^{+4.8}_{-4.1}$ & $15.6^{+4.5}_{-3.5}$ & $17.6^{+6.2}_{-5.3}$ & $11.2^{+2.7}_{-2.1}$ & $6.5^{+2.1}_{-1.7}$ \\
\\[-0.3cm]
\diskpbb\ & $T_{\rm{in}}$ & [keV] & $3.0^{+0.3}_{-0.6}$ & $1.6^{+0.1}_{-0.3}$ & $2.7^{+0.7}_{-0.4}$ & $2.4^{+0.7}_{-0.2}$ & $2.3^{+0.5}_{-0.2}$ & $3.0 \pm 0.7$ & $1.41^{+0.05}_{-0.06}$ & $1.34^{+0.04}_{-0.06}$ & $1.38^{+0.04}_{-0.05}$ & $1.18^{+0.17}_{-0.11}$ & $3.2^{+0.5}_{-0.6}$ \\
\\[-0.3cm]
& $p$ & & $0.579^{+0.005}_{-0.004}$ & $0.67 \pm 0.02$ & $0.59^{+0.02}_{-0.01}$ & $0.65^{+0.02}_{-0.03}$ & $0.68^{+0.04}_{-0.02}$ & $0.58 \pm 0.02$ & $0.75^{+0.09}_{-0.05}$ & $0.75^{+0.12}_{-0.07}$ & $0.70^{+0.06}_{-0.04}$ & $>0.72$ & $0.57 \pm 0.01$ \\
\\[-0.3cm]
& Norm & [$10^{-3}$] & $2.5^{+0.4}_{-0.8}$ & $102^{+144}_{-31}$ & $3.7^{+3.0}_{-2.2}$ & $6.1^{+2.7}_{-3.7}$ & $8.8^{+4.3}_{-3.3}$ & $1.9^{+1.4}_{-1.0}$ & $168^{+86}_{-45}$ & $198^{+118}_{-52}$ & $145^{+57}_{-32}$ & $303^{+379}_{-145}$ & $1.6^{+1.3}_{-0.6}$ \\
\\[-0.3cm]
\simpl\ & $\Gamma$\tmark[a] & & $3.53^{+0.13}_{-0.12}$ & -- & -- & -- & -- & -- & -- & -- & -- & -- & -- \\
\\[-0.3cm]
& $f_{\rm{scat}}$ & [\%] & $>38$ & $>42$ & $>37$ & $>48$ & $>58$ & $>27$ & $9^{+2}_{-3}$ & $<4$ & $<3$ & $>77$ & $>15$ \\
\\[-0.3cm]
\hline
\\[-0.2cm]
\chisq/DoF & & & 21696/21489 \\
\\[-0.3cm]
\hline
\hline
\\[-0.2cm]
\multicolumn{14}{c}{Magnetic Accretor Model: \tbabs $\times$ $($ \diskbb\ $+$ \diskpbb\ $+$ \cutoffpl\ $)$} \\
\\[-0.25cm]
\tbabs\ & $N_{\rm{H,int}}$\tmark[a] & [$10^{21}$ cm$^{-2}$] & $1.41^{+0.07}_{-0.03}$ & -- & -- & -- & -- & -- & -- & -- & -- & -- & -- \\
\\[-0.3cm]
\diskbb & $T_{\rm{in}}$ & [keV] & $0.32^{+0.01}_{-0.02}$ & $0.25 \pm 0.04$ & $0.26 \pm 0.03$ & $0.32^{+0.04}_{-0.03}$ & $0.32^{+0.04}_{-0.03}$ & $0.49^{+0.20}_{-0.07}$ & $0.27^{+0.03}_{-0.02}$ & $0.30^{+0.02}_{-0.03}$ & $0.26 \pm 0.03$ & $0.31 \pm 0.02$ & $0.34^{+0.03}_{-0.02}$ \\
\\[-0.3cm]
& Norm & & $8.0^{+1.9}_{-1.4}$ & $14.0^{+12.6}_{-7.0}$ & $25.3^{+16.1}_{-8.2}$ & $9.2^{+3.7}_{-2.4}$ & $9.3^{+3.7}_{-2.6}$ & $0.8^{+1.4}_{-0.7}$ & $21.5^{+8.5}_{-5.2}$ & $18.1^{+5.9}_{-4.1}$ & $21.4^{+10.2}_{-6.6}$ & $11.2^{+3.5}_{-2.4}$ & $5.6^{+2.7}_{-1.9}$ \\
\\[-0.3cm]
\diskpbb\ & $T_{\rm{in}}$ & [keV] & $2.7 \pm 0.2$ & $1.90 \pm 0.04$ & $2.6^{+0.5}_{-0.3}$ & $2.4 \pm 0.4$ & $2.2^{+0.4}_{-0.3}$ & $2.7^{+0.7}_{-0.5}$ & $1.47 \pm 0.05$ & $1.35^{+0.05}_{-0.04}$ & $1.39 \pm 0.04$ & $1.65 \pm 0.08$ & $2.8^{+0.7}_{-0.4}$ \\
\\[-0.3cm]
& $p$ & & $0.55 \pm 0.01$ & $0.63 \pm 0.01$ & $0.56 \pm 0.02$ & $0.64^{+0.13}_{-0.05}$ & $>0.66$ & $0.55 \pm 0.03$ & $0.72^{+0.06}_{-0.04}$ & $0.73^{+0.09}_{-0.05}$ & $0.69 \pm 0.04$ & $0.73^{+0.08}_{-0.05}$ & $0.54 \pm 0.01$ \\
\\[-0.3cm]
& Norm & [$10^{-3}$] & $2.1^{+0.7}_{-0.5}$ & $44^{+6}_{-5}$ & $2.3^{+2.0}_{-1.1}$ & $4.0^{+7.0}_{-2.2}$ & $8.1^{+14.9}_{-4.8}$ & $1.8^{+2.6}_{-1.0}$ & $131^{+45}_{-29}$ & $177^{+77}_{-45}$ & $136^{+43}_{-27}$ & $72^{+36}_{-20}$ & $1.8^{+1.5}_{-0.9}$ \\
\\[-0.3cm]
\cutoffpl\ & $\Gamma$ & & $0.59$\tmark[b] & -- & -- & -- & -- & -- & -- & -- & -- & -- & -- \\
\\[-0.3cm]
& $E_{\rm{fold}}$ & [keV] & $7.9$\tmark[b] & -- & -- & -- & -- & -- & -- & -- & -- & -- & -- \\
\\[-0.3cm]
& Norm & [$10^{-4}$] & $2.4 \pm 0.2$ & $2.3 \pm 0.1$ & $2.6^{+0.3}_{-0.4}$ & $2.5^{+0.2}_{-0.4}$ & $2.5^{+0.2}_{-0.3}$ & $2.3^{+0.3}_{-0.6}$ & $0.28 \pm 0.09$ & $<0.12$ & $<0.09$ & $2.1 \pm 0.2$ & $2.0^{+0.4}_{-0.9}$ \\
\\[-0.3cm]
\hline
\\[-0.2cm]
\chisq/DoF & & & 21705/21490 \\
\\[-0.3cm]
\hline
\hline
\\[-0.2cm]
$F^{\rm{obs}}_{0.3-1.0}$\tmark[c] & \multicolumn{2}{c}{\multirow{5}{*}{[$10^{-12}$\,\ergpcmsqps]}} & $1.15 \pm 0.02$ & $1.63 \pm 0.02$ & $1.15 \pm 0.06$ & $0.69 \pm 0.04$ & $0.61 \pm 0.03$ & $0.84 \pm 0.05$ & $1.34 \pm 0.03$ & $1.39 \pm 0.04$ & $1.37 \pm 0.04$ & $1.12 \pm 0.03$ & $1.12 \pm 0.03$ \\
\\[-0.3cm]
$F^{\rm{obs}}_{1.0-10.0}$\tmark[c] & & & $9.7 \pm 0.1$ & $19.2 \pm 0.2$ & $8.9 \pm 0.2$ & $7.1 \pm 0.1$ & $6.8 \pm 0.1$ & $8.3 \pm 0.2$ & $12.2 \pm 0.2$ & $11.1 \pm 0.2$ & $10.9 \pm 0.2$ & $13.1 \pm 0.2$ & $8.9 \pm 0.2$ \\
\\[-0.3cm]
$F^{\rm{obs}}_{10.0-40.0}$\tmark[c] & & & $3.5 \pm 0.1$ & $3.4 \pm 0.1$ & $3.4 \pm 0.1$ & $3.1 \pm 0.1$ & $3.1 \pm 0.1$ & $3.1 \pm 0.2$ & $0.55 \pm 0.08$ & $0.20 \pm 0.05$ & $0.19^{+0.05}_{-0.03}$ & $2.7 \pm 0.1$ & $2.9 \pm 0.2$ \\
\\[-0.3cm]
$F^{\rm{obs}}_{0.3-40.0}$\tmark[c] & & & $14.3 \pm 0.2$ & $24.3 \pm 0.2$ & $13.5 \pm 0.3$ & $10.9 \pm 0.2$ & $10.5 \pm 0.2$ & $12.2 \pm 0.3$ & $14.1 \pm 0.3$ & $12.7 \pm 0.3$  & $12.5 \pm 0.3$ & $16.9 \pm 0.3$ & $12.9 \pm 0.3$ \\
\\[-0.3cm]
\hline
\\[-0.2cm]
$L^{\rm{int}}_{0.3-40.0}$\tmark[d] & \multicolumn{2}{c}{[$10^{40}$\,\ergps]} & $2.55 \pm 0.04$ & $4.21 \pm 0.05$ & $2.43 \pm 0.06$ & $1.87 \pm 0.04$ & $1.78 \pm 0.04$ & $2.14 \pm 0.06$ & $2.56 \pm 0.06$ & $2.36 \pm 0.06$ &  $2.33 \pm 0.06$ & $2.92 \pm 0.05$ & $2.34 \pm 0.06$ \\
\\[-0.3cm]
\hline
\hline
\end{tabular}}
\label{tab_param}
\end{center}
$^{a}$ These parameters are globally free to vary, but are linked across all epochs. \\
$^{b}$ These parameters are fixed to the average values seen from the pulsed emission
from the currently known ULX pulsars, and are common for all epochs. \\
$^{c}$ The total observed flux in the full 0.3--40.0\,keV band, and the 0.3--1.0, 1.0--10.0
and 10.0--40.0\,keV sub-bands, respectively (consistent for both models). \\
$^{d}$ Absorption-corrected luminosity in the full 0.3--40.0\,keV band (consistent for
both models). These values assume isotropic emission, and may therefore be upper
limits. \\
\end{sidewaystable*}

\subsection{Luminosity vs Temperature}

In order to further explore the long-term behaviour of \hoix, we now fit the two
main continuum models discussed above (\ie\ for non-magnetic and magnetic
accretors) to the full set of broadband observations. Here, we follow the analysis
presented in \cite{Walton20} of the multi-epoch broadband data available for
another ULX, NGC\,1313 X-1. In short, we fit these two continuum models to all the
available broadband spectra simultaneously. In both cases, we assume that the neutral
absorption column is the same for all observations (\eg\ \citealt{Miller13ulx}), although
this is globally free to vary. For the magnetic accretor model, we again assume the same
spectral form for the accretion column as in Section \ref{sec_soft} (a \cutoffpl\ model
with $\Gamma = 0.59$ and $E_{\rm{cut}} = 7.9$\,keV, based on the average
spectral shape of the accretion columns in the known ULX pulsars). For the
non-magnetic accretor model with an up-scattering corona, we again use the \simpl\
model, and we assume a common photon index for all observations (the high-energy
spectra are very similar in eight out of the eleven broadband observations, and the one
case during the softer-state where this higher energy component is seen is also
consistent with having the same slope as in the other observations), although again
this is free to vary globally.

Both of these continuum models provide excellent global fits to the multi-epoch
broadband data; the non-magnetic accretor model gives a total \chisq/DoF =
21696/21489, while the magnetic accretor model gives a total \chisq/DoF =
21705/21490. The best-fit parameters for both models are given in Table
\ref{tab_param}.

\begin{figure}
\begin{center}
\hspace*{-0.2cm}
\rotatebox{0}{
{\includegraphics[width=240pt]{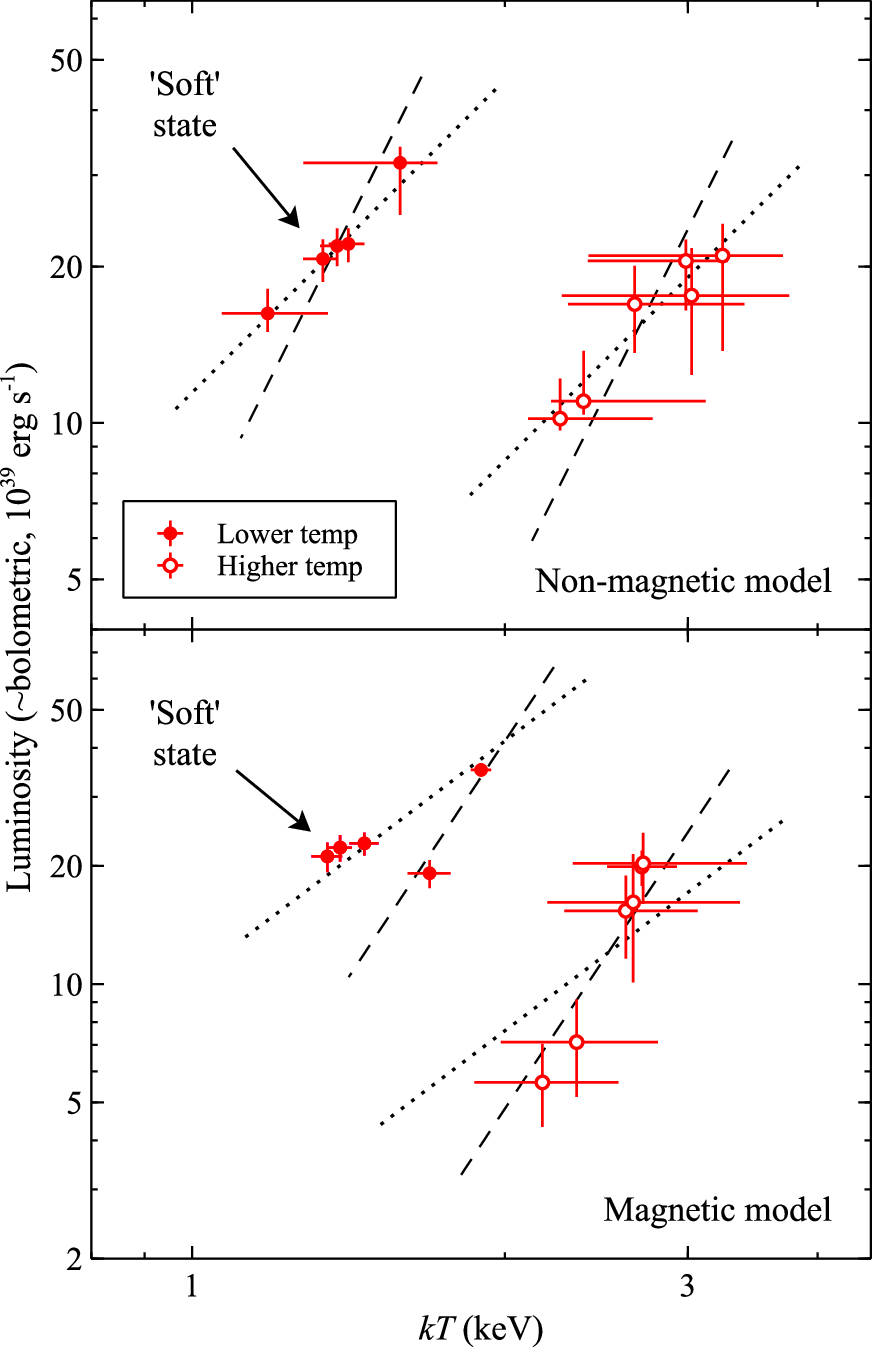}}
}
\end{center}
\vspace*{-0.3cm}
\caption{
0.001--100\,keV (\ie $\sim$bolometric) luminosity vs temperature for the \diskpbb\
component from our multi-epoch broadband spectral analysis of \hoix. Results for the
models assuming the accretor is non-magnetic and magnetic are shown in the top and
bottom panels, respectively. In the former case, evidence for two distinct
luminosity--temperature tracks are seen, both of which are consistent with either
$L \propto T^{4}$ (dashed lines) or $L \propto T^{2}$ (dotted lines), similar to the
results found for NGC\,1313 X-1 (\citealt{Walton20}). In the latter case, the situation is
less clear, with the soft state observations clearly separating themselves from the
other high-flux observations in the luminosity--temperature plane. The remaining
observations are consistent with there being two distinct luminosity--temperature
tracks that follow $L \propto T^{4}$, but with only two observations on the
high-flux/low-temperature track the presence of these two tracks cannot be well
established for the magnetic accretor model with the current data.
}
\label{fig_LT}
\end{figure}

Given the similarity of the behaviour in the archival data for \hoix\ and that seen from
NGC\,1313 X-1, we are particularly interested in the behaviour of the \diskpbb\ 
component included in both models in the luminosity--temperature plane. Following
\cite{Walton20}, we compute the luminosities of this component over a broad enough
band to be considered bolometric (0.001-100\,keV), and plot the
luminosity--temperature evolution for both models in Figure \ref{fig_LT}. For
comparison, in addition to the data we show example trends following $L \propto
T^{4}$ and $L \propto T^{2}$. The former scaling is expected for blackbody emission
from with a constant emitting area, and the latter is predicted by \cite{Watari00} for
an advection-dominated super-Eddington disc with a constant inner radius around a
black hole (the flatter relation comes about because some of the radiation emitted by
the disc becomes trapped within it and is carried over the event horizon instead of
escaping to infinity).

The two models show different behaviours for \hoix, relating in particular to the
new, soft state highlighted above. In the magnetic accretor model, the higher
temperature observations do show evidence of following a coherent
luminosity--temperature relation which appears to be slightly better described with $L
\propto T^{4}$ than with $L \propto T^{2}$. However, the lower temperature
observations (including the soft state) do not show a single, well-defined trend.
One could choose to connect the two lower-temperature observations in which the
high-energy emission is at its normal level (epochs 2 and 10), in which case these
data would roughly follow an $L \propto T^{4}$ trend with the soft state observations
(which cluster together) appearing as outliers.  Alternatively, one could choose to
connect the soft state observations with the brightest broadband observation to
date (epoch 2), in which case these data would roughly follow an $L \propto T^{2}$
trend with epoch 10 being an outlier. Given the significant difference in high-energy
flux seen in the soft state observations, the former may be the more natural scenario
in this case. Nevertheless, it is worth nothing that, while advection over a horizon is
not possible for neutron star accretors, if the accretor is magnetized with the disc
truncated at the magnetosphere it is still possible to have luminosity--temperature
relations that are flatter than $L \propto T^{4}$, as the radius of the magnetosphere is
generally expected to vary with accretion rate in an inverse manner (although there
may be some range of accretion rates where this is not the case; \citealt{Chashkina19}). 

In contrast, the non-magnetic accretor model does seem to show two well-defined
luminosity--temperature tracks, similar to the results seen for NGC\,1313 X-1
(\citealt{Walton20}). In this case, both the higher and lower temperature trends
seem to be slightly better described with $L \propto T^{2}$ than $L \propto T^{4}$,
but ultimately the data would appear to be consistent with both possibilities. The key
difference here is the use of the \simpl\ model to generate the highest energy
emission. In contrast to the use of the \cutoffpl\ model in the magnetic accretor
model, which is an additive component that is not formally coupled to the \diskpbb\
component of interest here, \simpl\ assumes that every photon that contributes to
the highest energy component was previously emitted by the disc. The bolometric
disc flux is calculated prior to the application of the \simpl\ component in our analysis.
As such, the properties inferred for the high-energy continuum have a more direct
impact on the properties inferred for the disc, resulting in the different
luminosity--temperature behaviour for the lower temperature observations when
compared to the magnetic accretor model, given the strong changes in the high-energy
flux during the soft state observations. 



The direct coupling between the \simpl\ and \diskpbb\ components described
above not only impacts the best-fit properties inferred for the disc, but also results
in stronger parameter degeneracies than for the magnetic accretor model, and thus
larger overall parameter uncertainties (the exceptions being the soft state
observations, where the high-energy continuum is sufficiently weak that the
degeneracies with the disc properties are minimised). However, with regards to
assessing the presence of the luminosity--temperature trends with this model,
these uncertainties are likely slightly misleading, as they are partially driven by
uncertainties in the high-energy photon index, which is strongly connected to the
inferred disc properties. This is treated as being common for all observations, and
as such changes in $\Gamma$ cause almost all the observations to change in
tandem. This degeneracy therefore expands the formal parameter uncertainties on
the disc properties for almost all the observations, but does not remove the need
for the relative variations between epochs that drive the overall presence of the
luminosity--temperature trends seen in Figure \ref{fig_LT}. Indeed, we have
confirmed that fixing $\Gamma$ to its best-fit value and to its $\pm$90\% parameter
bounds reduces the average fractional parameter uncertainties for both the luminosity
and the temperature measurements in all three cases, and that two positive
luminosity--temperature trends are still seen when pushing $\Gamma$ to these limits.

Again following the analysis presented in \cite{Walton20}, in order to alleviate any
concerns that the luminosity--temperature behaviour shown in Fig. \ref{fig_LT} may
be driven by the use of fluxes that are significantly extrapolated beyond the observed
bandpass, we also re-assess the luminosity--temperature behaviour for the \diskpbb\
component using fluxes calculated above 1\,keV instead. While there are necessarily
quantitative differences in the fluxes relating to this change in bandpass, for both
the models considered the same qualitative behaviour to that shown in Figure
\ref{fig_LT} is still seen with this alternative analysis. We have also investigated
relaxing the assumption that all epochs share a common column density, allowing this
to vary between epochs, and again the same behaviour as shown in Figure \ref{fig_LT}
is seen.

\section{Discussion}
\label{sec_dis}

We have conducted a series of five new broadband observations of the ULX \hoix,
combining \xmm\ and \nustar. Along with the broadband observations available in the
archive, combining \xmm, \suzaku\ and \nustar\ (epochs 1--6, reported in
\citealt{Walton14hoIX, Walton17hoIX, Luangtip16, Gurpide21}), we therefore have a
set of eleven broadband observations of \hoix. Compared to these archival datasets,
the first three of these new observations (epochs 7--9) show a new `soft' state in
which the hard X-ray flux, previously seen to be remarkably stable, has almost
completely collapsed (see Figure \ref{fig_spec_4states}). The final two new
observations (epochs 10 and 11) are broadly similar to the first two broadband
observations taken, and show a return to the same level of hard X-ray flux seen
previously (specifically, epochs 10 and 2 are similar, while epochs 11 and 1 are similar).
This recovery of the hard X-ray flux took $\lesssim$12 days.

\subsection{Luminosity--Temperature Behaviour}

The consistency of the high-energy flux seen in the archival \hoix\ datasets has also
been seen in the multi-epoch broadband data available for the ULX NGC\,1313 X-1
(\citealt{Walton20}). When the NGC\,1313 X-1 data were fit with what are now fairly
standard (albeit still phenomenological) ULX continuum models, combining
two thermal components that dominate below 10\,keV and an additional high-energy
continuum seen above $\sim$10\,keV (either an accretion column or an up-scattering
corona), the results for the hotter thermal component separated themselves into two
groups: one with higher temperatures and lower fluxes, and one with lower
temperatures and higher fluxes.  However, each of these groups exhibited its own
\textit{positive} luminosity--temperature trend (i.e.  luminosity increasing with
temperature).  As the archival broadband observations of \hoix\ only include one
observation that would correspond to the low-temperature/high-flux group seen in
NGC\,1313 X-1, the original purpose of the new observations presented here was to
investigate whether the hotter thermal component in \hoix\ also showed the same
complex luminosity--temperature behaviour as NGC\,1313 X-1, given the other
similarities in their broadband spectral evolution.

To test this, we performed a similar,  multi-epoch spectral analysis to
\cite{Walton20} on the updated broadband data for \hoix. Based on these fits, there
is some evidence that \hoix\ may show the same strange luminosity--temperature
behaviour as seen in NGC\,1313 X-1, though the results for \hoix\ are model dependent.
The case is strongest for the non-magnetic accretor model; the results for the hotter
thermal component (treated with the \diskpbb\ model) show two clear
luminosity--temperature tracks, with all the lower-temperature epochs seeming to
follow a coherent trend (including the new soft-state observations). The results for the
magnetic accretor model are more ambiguous; here the lower-temperature
observations do not show a single coherent trend. However, if the new soft-state
observations are treated as being distinct to the other lower-temperature observations
(\ie those where there is significant hard X-ray flux), which may not be unreasonable
given the differences at higher energies, then one can interpret the remaining
observations as being consistent with showing two distinct luminosity--temperature
tracks similar to NGC\,1313 X-1 (although there are then only two observations on the
higher-flux/lower-temperature track). Were this to be the case, the data would seem to
indicate a preference for $L \propto T^{4}$ trends.

The cause of this behaviour is difficult to explain, particularly given that
the high-energy flux seen in the broadband NGC\,1313 X-1 data (and the majority of
the broadband \hoix\ data) is so stable and that a relatively sharp transition between
the two trends is needed in the luminosity--temperature plane. \cite{Walton20}
consider a variety of different possible scenarios, including:

\begin{itemize}

\item Changes in the degree of geometric collimation/beaming\footnote{Note that
while this is often referred to as `beaming' in the literature, \eg\ \cite{King08}, it is
a geometric effect that is distinct from relativistic beaming.} experienced by the
emission from the innermost regions of the disc, i.e. the regions interior to the onset
of the funnel-like geometry -- which results in the collimation -- expected for the
inner regions of super-Eddington accretion discs.

\item The inner disc becoming completely blocked from view, such that only the outer
disc/wind is seen.

\item The inner disc being covered by a scattering shroud, potentially related to an
ionised wind (since extreme outflows are known to be present in NGC\,1313 X-1;
\citealt{Pinto16nat, Pinto20, Walton16ufo}), resulting in a diminished (but non-zero)
flux from these regions of the disc reaching the observer.

\item Changes in atmospheric effects in the disc (i.e. its colour correction factor)
and/or the amount of down-scattering in the wind.

\item Emission from distinct (e.g. radially segregated) regions of the accretion flow
having different variability properties such that they each dominate the total emission
at different times. These regions would likely be determined by the key radii expected
for super-Eddington accretion, \ie the inner disc radius (set by the innermost stable
circular orbit for a black hole, the neutron star surface for a non-magnetised neutron
star, or the magnetospheric radius for a magnetic neutron star) and the `spherisation'
radius.

\end{itemize}

However, in the end none of these possibilities were considered to offer an entirely
satisfactory explanation for the broadband data for NGC\,1313 X-1. We do not repeat
that discussion in detail here, and instead refer the reader to that work for a more
in-depth consideration of these possibilities. However, to briefly summarise the main
issues, scenarios invoking geometric changes to produce the two
luminosity--temperature tracks (\ie differences in beaming, occultation) struggled to
explain the lack of corresponding variability at the highest energies, as regardless of
whether this emission is associated with a scattering corona or an accretion column
the general expectation is that this emitting region should be central and compact,
and so should respond similarly to anything impacting the emission from the inner
disc. A number of the other scenarios discussed would have implied bi-modal
behaviour (or at least sharp transitions) in quantities that would likely be expected
to vary smoothly with accretion rate (\eg\ the amount of beaming, the colour
correction factor) to produce two distinct luminosity--temperature tracks. 

The final possibility, invoking multiple different regions in a complex, super-Eddington
disc is likely the most compelling, but here it is then also necessary to explain the
further presence of the second, cooler thermal component at $\sim$0.3\,keV required
by the broadband data (such that there are at least three distinct regions of the
accretion flow that manifest themselves in the data, in addition to the high-energy
corona/accretion column). For a magnetised neutron star accretor, one notable
possibility is that the two luminosity--temperature tracks correspond to distinct
thermal contributions from the super-Eddington disc interior to \rsp, and the
accretion curtain that connects the disc at \rmag\ to the accretion columns, which
\cite{Mushtukov17} suggest should be optically thick. However, it is not entirely clear
why the intermediate temperature regions (corresponding to the cooler of the
luminosity--temperature tracks seen from the \diskpbb\ component) would dominate
the total emission on some occasions but not on others.

While the high-energy flux from \hoix\ has now been seen to vary significantly in
these new observations, there are still at least some observations on the
higher-flux/lower-temperature track that have equivalent high-energy flux to the
observations on the higher-temperature/lower-flux track. As such, all the issues
explaining the two luminosity--temperature tracks with geometric changes
discussed in \cite{Walton20} for NGC\,1313 X-1 would still also hold for \hoix, despite
this newly observed variability.

Ultimately, though, given the more model-dependent nature of the results found here,
further broadband observations will be required to unambiguously confirm that \hoix\ is
truly exhibiting the same luminosity--temperature behaviour as seen in NGC\,1313 X-1.
In particular, these observations should target periods of high overall flux where there
is also significant hard X-ray flux as well (i.e. the flux above $\sim$10\,keV is at its
`normal' level) in order to investigate whether these observations really show a
coherent luminosity--temperature trend when the magnetic accretor model is applied.

\subsection{The Nature of the Soft State}
\label{sec_soft_dis}

We focus the remainder of our discussion on the nature of the soft state, and in
particular the processes that could be responsible for the collapse in the observed
hard X-ray flux.  We consider four main possibilities:

\begin{itemize}

\item Obscuration of the inner regions of the accretion flow by regions of the
super-Eddington disc/wind residing at larger radii

\item The formation of a `scattersphere' within the funnel-like geometry
expected for a super-Eddington disc/wind combination

\item A transition into the propeller regime (which would specifically require
a magnetized neutron star as the central accretor)

\item Collapse/destruction of a Compton-scattering corona

\end{itemize}

In brief, it is not clear that either of the scattersphere or propeller scenarios offer
viable explanations for the observed soft state when these data are considered in
the context of the multi-epoch broadband dataset now available for \hoix. While
obscuration by a super-Eddington disc/wind could be viable under certain
conditions, our current view is that a scenario involving the destruction of a
Compton-scattering corona may be the most plausible interpretation (though we
stress that arriving at a firm conclusion is challenging with the currently available
data). More detailed discussions of the scenarios we consider to be potentially
viable are presented in the following subsections for the interested reader, while
the others are discussed further in Appendix \ref{app_soft}.

\subsubsection{Obscuration}
\label{sec_obscur}

One possibility is that the during the first three of the new observations the innermost
regions were blocked from view, preventing us from seeing the highest energy emitting
regions. Given the strong remaining thermal emission, the likely obscuring medium
would be regions of a geometrically thick, super-Eddington accretion disc and/or its
wind at larger radii. This is conceptually similar to one of the scenarios considered
above for the complex luminosity--temperature behaviour, although such an
obscuration-based scenario does not ultimately seem a plausible explanation for that
phenomenology. It is also analogous to the scenario often invoked to explain ULXs that
transition into the `ultrasoft' regime (now referred to as ultraluminous supersoft
sources, or ULSs; e.g. \citealt{Urquhart16, Pinto17, Dai21}), although we stress that the
`soft' state seen here differs markedly to these ULSs (the observed thermal emission
is still much hotter than the $\sim$0.1\,keV emission that characterises ULSs, and most
ULS observations exhibit luminosities around the $\sim$10$^{39}$\,\ergps\ level while
the soft state observations show luminosities of $\sim$10$^{40}$\,\ergps).

In principle, this would be possible regardless of whether the highest energy emission
originates in an up-scattering corona or an accretion column since, as noted above,
both would be expected to be compact and centrally located. However, based on the
spectral results from our multi-epoch analysis, this scenario would likely be more
plausible for the magnetic accretor model. Obscuration of the inner regions would
likely also be expected to result in an accompanying sharp drop in the temperature
seen from the thermal emission, as the innermost regions of the disc would likely also
be obscured in addition to the corona/accretion column. When compared to the
other high-flux observations (epochs 2 and 10), evidence for such a drop is only
really seen with the magnetic accretor model, while the non-magnetic accretor model
seems to show a smooth, coherent luminosity--temperature trend.

Even then, though, this emitting region would need to be only partially covered
during epoch 7, as while it is still much weaker, the high-energy emission component
seen during most other observations is still detected during this epoch. It is unclear
how plausible such a partially covering scenario is. Furthermore, the scale-height of
the disc/wind is generally expected to increase with increasing accretion rate (\eg\
\citealt{King08}). For any source where we typically can view the inner regions, one
would therefore generally expect the onset of such occultation to occur at the
highest accretion rates observed. However, the brightest observation (epoch 2) is
brighter than all of the soft state observations (both in terms of total flux and the
\diskpbb\ flux specifically), which would naively appear to imply a higher accretion
rate for epoch 2. In order for this scenario to be plausible, the soft state observations
would likely have to be intrinsically brighter than even epoch 2, and only appear
fainter because the most luminous inner regions are blocked from
view.\footnote{Although we do not report on this analysis in full here, we note that we
have also inspected the integrated soft state data from the \xmm\ Reflection Grating
Spectrometer (RGS; \citealt{XMM_RGS}) to check for any ionised absorption that
may further support this scenario. We do not find any notable evidence for such
absorption, but stress that these data are not particularly sensitive; the individual
\xmm\ exposures taken during this state are short, and the situation is further
exacerbated by the technical issue that occured during epoch 9, as this also resulted
in the loss of the RGS data from that epoch too, meaning there is only $\sim$43\,ks
good RGS exposure covering the soft state. Deeper exposures targeting this state
will be required to meaningfully perform meaningful searches for such absorption.}
This would have further implications for the complex luminosity--temperature
behaviour observed; if the soft-state is what the source looks like when the
innermost regions are obscured from view, then this would offer a further argument
against the presence of the two `normal' luminosity--temperature tracks being
related to similar obscuration.

\subsubsection{Collapsing/Disappearing Corona}

Should the high-energy flux come from an up-scattering corona, the observed
variability could instead be related to intrinsic changes in this structure. The fraction
of disc photons scattered up into the high-energy component changes significantly
in this model, ranging from upwards of $\sim$75\% (epoch 10) to less than 3\%
(epoch 9). This would indicate that the corona has essentially collapsed during the
soft state observations, before re-forming by the time of epoch 10 ($\sim$12 days
after epoch 9, the last soft-state observation). It is particularly interesting that, despite
this very large change, the intrinsic properties inferred for the \diskpbb\ component
during all of the high-flux observations (including the soft state observations) show
a single, coherent trend. Again, we note that there is no reason we are aware of that
this should occur by coincidence.

An interesting comparison here may be the strange recent behaviour seen in the
active galactic nucleus 1ES\,1927+654. Although a long-known AGN
(\citealt{Boller03_1927, Gallo13_1927}), in late 2017 this source underwent a transient
outburst event during which the hard X-ray corona was destroyed sometime in mid
2018, leaving an extremely soft, vaguely thermal spectrum with a temperature of
$\sim$100\,eV and essentially no flux above $\sim$a few keV. Continued monitoring
showed that the corona had begun to re-form by late 2018 and the source eventually
reached X-ray luminosities significantly brighter than its pre-outburst state (although
it has since returned to comparable levels; \citealt{Ricci20, Masterson22}). The cause
for this outburst behaviour is not known, but it is speculated that this may have been
triggered by a rare example of a tidal disruption event (TDE) in an existing AGN. TDEs
are often thought to result in super-Eddington accretion, and indeed, at its peak the
X-ray flux in just the 0.3--10.0\,keV band alone reached the Eddington limit for the
$10^{6}$\,\msun\ black hole this source is expected to host (\citealt{Ricci20}), implying
that the overall accretion rates were at times super-Eddington. Interestingly, during its
peak fluxes the re-formed corona in 1ES\,1927+654 exhibited a very steep spectrum
($\Gamma \sim 3$), somewhat similar to the coronal spectrum inferred here. The
timescale over which the corona was re-created in 1ES\,1927+654 ($\sim$100 days)
would easily fit in the $\sim$12-day period between epochs 9 and 10 when scaled
from a $\sim$10$^{6}$\,\msun\ black hole to e.g. a 10\,\msun\ black hole, although the
duration for which the corona was absent in \hoix\ would appear to be much longer
than for 1ES\,1927+654 (again when scaled by the mass).

The exact process by which the corona was destroyed in 1ES\,1927+654 is not well
known, but given the long-term variability seen from \hoix\ (Figure \ref{fig_longlc}) we
cannot invoke anything conceptually similar to a TDE here. If these two collapses in
flux really are similar then it would therefore seem more likely that they are somehow
related to the high/super-Eddington rates of accretion in both sources, as opposed to
dynamical processes relating to a TDE event. \cite{Ricci20} speculate that the
destruction of the corona in 1ES\,1927+654 is related to the destruction of the
innermost accretion disc, which would normally support the X-ray corona (perhaps
magnetically). However, we would expect such a sudden change in the disc properties
to manifest itself in the luminosity--temperature plane. For \hoix\ we do not see any
evidence for this when the high-energy emission is modelled as an up-scattering
corona. One could perhaps argue that the inner disc was initially destroyed along with
the corona in \hoix, but had since re-formed prior to epoch 7, as this may have to
occur prior to the main recovery of the corona. However, the detection of the weak
high-energy flux during epoch 7 may suggest that we are catching the end of the
destruction process, and again there are no major differences between the disc
properties during this epoch and the other soft state observations. This may instead
imply that there is a way of destroying the corona without significantly impacting the
disc. Alternatively, the re-formation of the corona may just not be an entirely smooth
process, with epoch 7 representing something like an initial, failed attempt before the
more sustained restoration occurred later on (between epochs 9 and 10).

One speculative possibility is that the collapse of the corona is somehow related to
ballistic jet ejections. During their outbursts, `normal' low-mass X-ray binaries are
known to launch transient, ballistic jet ejecta as they approach luminosities close to
their Eddington limits (\eg\ \citealt{Corbel02, Tomsick03, Corbel05, Steiner12h1743}).
A small number of ULXs are also known to exhibit similar transient ejections
(\citealt{Middleton13nat, Cseh14, Cseh15}), and \hoix\ would appear to be among this
number.\footnote{Infrared (IR) emission consistent with a jet was originally reported by
\cite{Dudik16}, and then further investigation of the full, multi-epoch \spitzer\ data
showed that this corresponded to a brief, transient period of IR activity (\citealt{Lau19}).
Recent constraints on the presence of a compact radio source associated with \hoix\
further confirm that the jet activity must be transient (\citealt{Berghea20}).} If these
ejection events can `clear out' the corona (at least on occasion), then this may offer a
means to temporarily remove the corona for some period of time after they occur. It is
worth noting that these transient jet events are often accompanied by strong X-ray
flares (\eg\ \citealt{McClintock09, Steiner11, Walton17v404}), and there are no obvious
events like this in the \swift\ monitoring immediately prior to the soft state
observations. However, the flares associated with these jet ejection events typically
reach $\sim$Eddington or mildly super-Eddington luminosities. It is therefore not
clear whether similar flares would be detectable for a source already radiating at
highly super-Eddington levels. Furthermore, these flares/flaring periods can be very
brief, and so even if flaring to even higher luminosities should have occurred, this
could easily have been missed by the $\sim$weekly \swift\ monitoring cadence.

Given all of these potential connections, and the coherent luminosity--temperature
relation implied by this model for all the high-flux observations, we speculate that a
scenario along these lines (whereby an up-scattering corona is somehow physically
destroyed) may be the most plausible explanation for this new soft state in \hoix.
However, we stress that an obscuration-related scenario cannot be excluded under
the right circumstances, and that this does not shed any further light on the fact that
there seem to be two distinct luminosity--temperature trends seen from the hotter
thermal component in this spectral model.

\section{\textit{HEX-P} Simulations}
\label{sec_hexp}

Although we may not have a satisfactory picture for the spectral evolution exhibited
by \hoix\ at the current time, neither in terms of observationally establishing whether
\hoix\ really exhibits the two distinct luminosity--temperature tracks similar to
NGC\,1313 X-1 nor understanding the nature of the new `soft' state revealed here,
addressing these issues is likely to play a key role in establishing our detailed
understanding of super-Eddington accretion. Such work requires continued
observations to build up a larger multi-epoch broadband dataset, particularly
targeting the `normal' high-flux state. For \hoix\ there is plenty of scope to explore
this issue further with additional broadband observations with our current facilities
(i.e. further coordinated observations with, e.g., \xmm\ and \nustar). Assuming that
similar behaviour to NGC\,1313 X-1 really is present in \hoix\ as well, as is tentatively
hinted by the current broadband observations, going on to further establish whether
this is common among the ULX population will also be of significant importance.
However, building the multi-epoch broadband datasets necessary to investigate this
for a population of sources would require a vast amount of coordinated exposure
with our current facilities.

\begin{figure}
\begin{center}
\hspace*{-0.2cm}
\rotatebox{0}{
{\includegraphics[width=240pt]{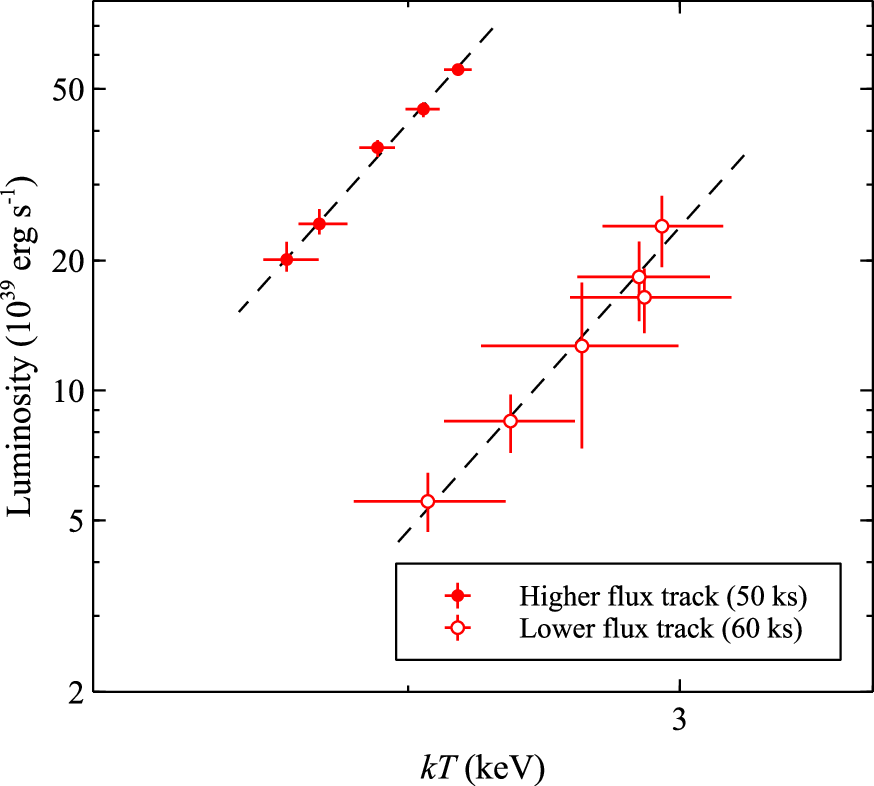}}
}
\end{center}
\vspace*{-0.3cm}
\caption{
Simulated \textit{HEX-P} luminosity--temperature constraints for the \diskpbb\ 
component. For simplicity, we focus only on the magnetic accretor model here, and
we also explicitly assume that \hoix\ does exhibit two distinct luminosity--temperature
tracks that each follow $L \propto T^{4}$ when the source has a `normal' hard X-ray
flux (\i.e. the hard X-ray flux seen in all observations apart from the new soft state
discovered here), similar to the results seen from NGC\,1313 X-1. In total 11
\textit{HEX-P} observations are simulated (matching the number of broadband
observations currently in the archive for \hoix), five in the higher-flux/lower temperature
track (50\,ks exposurs) and six in the lower-flux/higher-temperature track (60\,ks
exposures), with a total exposure of $\sim$600\,ks. The luminosity--temperature
constraints from these \textit{HEX-P} observations would be either equivalent or
superior to the current constraints that combine $\sim$\totcurrent\ of observing time
from \xmm, \suzaku\ and \nustar.
}
\label{fig_HEXP}
\end{figure}

The recently-proposed \textit{High Energy X-ray Probe} (\hexp; \citealt{HEXP_2024})
will offer simultaneous broadband coverage (bandpass of $\sim$0.3--80.0\,keV)
with unprecedented hard X-ray sensitivity (at least a factor of four better than \nustar\
over their common energy range), and would thus be ideally suited for such work. In
order to illustrate this, we simulate a multi-epoch set of \hexp\ observations of \hoix,
assuming for illustration that the \diskpbb\ component in this source does indeed
exhibit two distinct luminosity--temperature trends, each of which follows $L \propto
T^{4}$. For simplicity, we focus on the magnetic accretor model for these simulations,
and we simulate the same number and make-up of \hexp\ observations as the overall
number of broadband observations as are currently available: six in the
lower-flux/higher-temperature group and five in the higher-flux/lower-temperature
group (11 in total). The lower flux/higher temperature simulations span temperatures
of 2--3\,keV, while the higher flux/lower temperature simulations span temperatures
of 1.65--2.15\,keV, while typical parameters for the cooler \diskbb\ component were
adopted based on the results presented earlier, and the hard X-ray flux (10--40\,keV,
which sets the normalisation of the highest energy component) was assumed to be
$3 \times 10^{-12}$\,\ergpcmsqps. For each of the simulated spectra, the magnetic
accretor model was then re-fit to the data with all the key parameters free to vary;
to be conservative we treated each of the simulated spectra independently (i.e. we
did not attempt to fit them together, linking parameters between them).

The results for the luminosity--temperature plane for the \diskpbb\ component from
these simulations are shown in Figure \ref{fig_HEXP}, assuming \hexphighTexp\,ks
exposures for the lower-flux/higher-temperature observations and \hexplowTexp\,ks
exposures for the higher-flux/lower-temperature observations, resulting in an overall
exposure of $\sim$\hexptotexp\,ks for the total set of observations. This is to be
compared against the total observational investment of \totcurrent\ that went into
producing the results presented in Figure \ref{fig_LT}. Either equivalently good or
superior results would be provided by \hexp\ with a significantly smaller overall
investment, and critically with no reliance on coordinated observations, demonstrating
the impact that \hexp\ would have on our understanding of the broadband spectral
evolution seen from ULXs. While the simulations presented here are primarily set up to
be relevant to \hoix\ (though they may also therefore be relevant to NGC\,1313 X-1;
\citealt{Walton20}), there is of course significant interest in determining the evolution
of the different ULX spectral components more generally (\eg\ \citealt{Gurpide21,
Robba21, Barra22, Barra24}). A further exploration of the ULX science that would be
possible with \textit{HEX-P} is presented in \cite{Bachetti23_hexp}.


\section{Summary and Conclusions}
\label{sec_conc}

We have presented results from a series of five new coordinated \xmm+\nustar\ X-ray
observations program on the ULX \hoix\ performed in late 2020 (bringing the total
number of broadband observations with high-energy coverage from \nustar\ to 11),
focusing on the broadband ($\sim$0.3--40.0\,keV) spectral evolution exhibited by
this source. While \hoix\ has previously shown evidence for remarkably stable
high-energy emission above $\sim$10--15\,keV (despite clear variations at lower
energies), the first three of these new observations reveal a new `soft' state in which
the observed high-energy flux has collapsed. The last two observations show a
recovery in the hard X-ray flux, which returns to its previously stable level. We consider
a variety of different possible scenarios for this sudden collapse in high-energy flux,
allowing for the origin of the high-energy emission to be either a Compton
up-scattering corona around a non-magnetic accretor or an accretion column onto a
magnetized accretor (using the known ULX pulsars as a template), as the nature of the
accretor in \hoix\ currently remains uncertain. These scenarios include obscuration by
the geometrically thick super-Eddington accretion disc expected in this system
(possible for both the corona and accretion column origins for the high-energy
emission), a transition into the propeller regime (should the high-energy emission
come from an accretion column) and the destruction and re-formation of a corona
analogous to the event recently seen from the AGN 1ES\,1927+654. Our current view
is that the latter may be the most plausible explanation, though a definitive conclusion
is challenging with the available data.

The interpretation of the high-energy emission in \hoix\ has an important impact on
the inferred behaviour of the thermal component assumed to be associated with the
innermost accretion disc (the \diskpbb\ component in the spectral models utilized
here) in the luminosity--temperature plane. Should this arise in a corona, the
multi-epoch data for this component appear to trace out two distinct
luminosity--temperature tracks similar to the results seen from the ULX NGC\,1313
X-1: a higher-flux/lower-temperature track and a lower-flux/higher-temperature track,
each of which shows a positive correlation between luminosity and temperature.
Should the high-energy emission instead arise from an accretion column the behaviour
is more complex, with the new soft-state observations separating themselves from the
other high-flux observations that have `normal' levels of high-energy emission. It could
still be the case that two distinct luminosity--temperature tracks (similar to NGC\,1313
X-1) are seen outside of this new soft-state, but with only two such epochs currently
available this remains highly speculative. Understanding the nature of this soft-state
and the luminosity--temperature behaviour of the accretion disc in \hoix\ are closely
linked, but further broadband observations of \hoix\ will be required to shed more light
on these issues.

\section*{ACKNOWLEDGEMENTS}

The authors would like to thank the reviewer for their helpful feedback, which helped
to improve the final version of the manuscript.
DJW, TPR and WNA acknowledge support from the Science and Technology Facilities
Council (STFC; grant codes ST/Y001060/1, ST/X001075/1 and ST/Y001982/1,
respectively). CP is supported by PRIN MUR SEAWIND -- European Union --
NextGenerationEU.
RS acknowledges support from the University of the Chinese Academy of Sciences
during part of this work, as well as support from the INAF grant number 1.05.23.04.04.
This research has made use of data obtained with \nustar, a project led by Caltech,
funded by NASA and managed by NASA/JPL, and has utilized the \nustardas\ software
package, jointly developed by the ASDC (Italy) and Caltech (USA).
This research has also made use of data obtained with \xmm, an ESA science mission
with instruments and contributions directly funded by ESA Member States. 


\section*{Data Availability}

The raw observational data underlying this article are all publicly available from ESA's
\textit{XMM-Newton} Science
Archive\footnote{https://www.cosmos.esa.int/web/xmm-newton/xsa} and NASA's
HEASARC archive\footnote{https://heasarc.gsfc.nasa.gov/}.

\bibliographystyle{/Users/dwalton/papers/mnras}

\bibliography{/Users/dwalton/papers/references}

\appendix

\section{Pulsation Searches}
\label{app_pulse}

In addition to the spectral analysis presented in the main body of the paper, we have
also undertaken pulsation searches for all of the \xmm\ and \nustar\ observations of
\hoix\ presented here, analysing the data from each epoch independently (we focus on
these observatories as they have detectors with sufficient temporal resolution to
detect the spin periods of $\sim$1\,s or less seen in most ULX pulsars). To facilitate
these searches, all photon arrival times were transferred to the solar barycenter using
the DE200 solar ephemeris.

To account for the high spin up rates commonly found in PULXs, we ran an
accelerated Fourier-domain search \citep{PRESTOaccel}, implemented in the tool
\texttt{HENaccelsearch} included in the HENDRICS software package
(\citealt{bachettiHENDRICSHighENergy2018}). For \nustar\ observations the maximum
frequency investigated was 1000 Hz, while for \xmm\ observations it was between
$\sim$7 and 200 Hz, depending on whether the observation was taken in full frame or
small window mode. \nustar\ events from FPMA and FPMB were merged into a single
event list for this analysis in order to improve the signal-to-noise ratio. Events outside
good time intervals were rejected.

Internally, \texttt{HENaccelsearch} substitutes the light curve outside GTIs with the
mean of the two adjacent GTIs. To overcome the loss of sensitivity of the power
density spectrum (PDS) close to the borders of the spectral bins, we used
``interbinning", a way to interpolate the signal between two bins using the signal in
the adjacent bins \citep{vanderklis89}. This search technique produces many false
positives due to the alteration of the white noise level of the PDS, all of which need to
be confirmed with independent methods. We found $\sim$100 candidates over the
whole frequency band. Most of them were at very low frequencies, clustered around
specific frequencies and their harmonics which could be traced to beats of the
orbital period. A few turned out to be clear artifacts (like sub-harmonics of the
sampling frequency, recognizable by the step-function shape of the folded profile)
and were rejected.

We then analysed the $f-\dot{f}$ parameter space around all remaining candidate
frequencies with the $Z^2_1$ (i.e. a Rayleigh test). To perform this analysis, we used
the tool \texttt{HENzsearch} with the ``fast" option; this optimizes the search in the
$\dot{f}$ space by pre-binning the photons in phase and computing the $Z^2_n$
statistic using these bins instead of the individual photons, a technique referred to as
the quasi-fast folding algorithm (\citealt{STINGRAY, STINGRAYv2}).


Most of the candidates were not confirmed as significant by the $Z^2_1$ search,
which is due to the altered statistical properties of interbinning. Several candidates
at high power ($Z^2_1 \gtrsim 50$) appeared, which would normally be considered
detections, but given the high number of trial frequencies and frequency derivatives
spanned by the acceleration search process, the interbinning and the interpolation
outside of GTIs, it is difficult to quantify exactly the number of trials involved, and in
turn, evaluate their significance. Moreover, none of the candidates showed up in
more than one dataset, or in both of the paired, quasi-simultaneous \xmm\ and
\nustar\ observations, which makes it unlikely they represent the rotation of a neutron
star. As such, we do not report any significant detection of pulsations from these
observations, and adopt the position throughout the main paper that the nature of
the accretor in \hoix\ remains unknown.

\section{Further Discussion of the Nature of the Soft State}
\label{app_soft}

As noted in Section \ref{sec_soft_dis},  though these were ultimately deemed
unlikely to be viable explanations,  here we provide further discussion of the two
remaining scenarios considered for the soft state in \hoix, providing our reasoning
as to why we consider them unlikely.

\subsection{Formation of a `Scattersphere'}

A possibility somewhat related to the obscuration scenario (Section
\ref{sec_obscur}) is that we are typically viewing \hoix\ down the funnel formed by
a super-Eddington disc/wind (as may be expected given its typically hard spectral
shape; \citealt{Sutton13uls}), but that during the soft state the optical depth of the
material inside the funnel itself becomes optically-thick to electron scattering, with
the resulting `scattersphere' blocking our view of the inner regions (instead of the
outer disc/wind). \cite{Narayan17} suggest this can result in a significant softening
of the osberved spectrum (see also \citealt{Kawashima12}), particularly if the
scattersphere lies at a large distance from the accretor where temperatures are
cooler. However, this scenario would be expected to produce a significantly
enhanced soft X-ray flux during the soft state relative to the other high-flux epochs.
The formation of such a scattersphere may require an increase in accretion rate
through the disc, but even if it can be produced by other changes in the disc
structure while \mdot\ remains $\sim$constant, all of the hidden high-energy flux
should now be down-scattered to the temperature of the scattersphere and
emerge at these energies instead. Such an enhancement in soft X-ray flux is not
obviously seen though (see Table \ref{tab_param}).

\subsection{Propeller Transition}

Should \hoix\ be powered by a magnetised neutron star, such that the highest energy
flux is associated with emission from one or more accretion columns, another
interesting possibility is that the disappearance of the high-energy flux is related to a
transition to the propeller regime\footnote{3D general-relativistic
magneto-hydrodynamic simulations have shown that even when a source is in the
propeller regime, occasionally it is still possible for small amounts of material to
penetrate down to the neutron star owing to the turbulent nature of the accreting
material outside of the magnetosphere, even though significant, continuous accretion
rates are prevented (\citealt{Parfrey17b}). As such, this small amount of high-energy
flux may not be an issue for this possibility}. The propeller regime corresponds to a
state in which the material at the magnetosphere is not rotating fast enough to couple
to the rotating magnetic field of the central object, meaning that the inflow of material
through the magnetosphere is not possible (\eg\ \citealt{Illarionov75}). Whether a
source is in the propeller regime thus depends on whether the magnetospheric radius
(\rmag) is inside or outside the `co-rotation' radius (\rco; the radius at which the
orbital motion in the disc matches the rotation period of the central object). If
\rmag\ $<$ \rco\ then the material is rotating fast enough and accretion can proceed
down to the neutron star surface, while if \rmag\ $>$ \rco\ then the material is not
rotating fast enough and accretion through the magnetosphere is halted. 

Since \rmag\ is generally expected to depend on the accretion rate (\eg\
\citealt{Cui97}), it is possible for sources to transition to and from the propeller regime.
Typically, transitions to the propeller regime are expected to be associated with a
catastrophic drop in X-ray flux, corresponding to the sudden cessation of accretion
through the magnetosphere (\eg\ \citealt{Cui97, Asai13, Tsygankov16prop,
Lutovinov19}). However, if there is substantial X-ray emission from a super-Eddington
accretion disc outside of the magnetosphere then it may be possible to sustain a high
overall observed flux, with the catastrophic drop limited to the energies at which the
emission from within \rmag\ dominates. Indeed, phase-resolved spectral decomposition
of the known ULX pulsars implies there is a significant contribution from non-pulsed
thermal emission in addition to the pulsed component associated with the accretion
column (\citealt{Walton18p13, Walton18ulxBB}). Furthermore, in some of the ULX
pulsars the pulsations are seen to come and go without large changes in
soft X-ray flux (\eg\ \citealt{Israel17}), which could potentially be due to the
accretion column shutting off after a propeller transition (resulting in a lack of
pulsations) while the X-ray-emitting super-Eddington disc outside of \rmag\ remains
(see also \citealt{Middleton23}).

A significant issue here, however, is that propeller transitions are usually expected to
be seen at low accretion rates as they are generally thought to be driven by changes
in accretion rate moving \rmag\ outside \rco, and the collapse of the high energy flux
in \hoix\ is actually seen during relatively bright observations. In order for this to be
the case, we would require that the observations corresponding to the
lower-flux/higher-temperature group in Fig. \ref{fig_LT} actually have higher accretion
rates than the rest of the broadband observations. In turn, this would require that the
complex luminosity--temperature behaviour is caused by something along the lines
of the scattering wind scenario described above (the sort of evolution shown in the
right-hand panel from Fig. 7 in \citealt{Walton20}). However,  we note again that the
similarity of the highest energy flux between the lower-flux/higher-temperature
observations and the `normal' higher-flux/lower-temperature observations is very
difficult to explain in this scenario, as the emission from the accretion columns must
come from even smaller scales than the inner disc, and so any screen that results in a
suppression of the observed flux from the inner disc should realistically also result in
a suppression of the observed flux from the accretion columns as well. 

Alternatively, given that what ultimately needs to occur for a propeller transition is a
switch from \rmag\ $<$ \rco\ to \rmag\ $>$ \rco, a propeller transition could
potentially be driven by a change in \rco\ instead of \rmag\ (\citealt{Middleton23}).
Indeed, in the model of super-Eddington accretion onto a magnetised neutron star
constructed by \cite{Chashkina19}, over a limited range of accretion rates \rmag\
actually becomes essentially independent of the accretion rate\footnote{This occurs
when the disc is dominated by radiation pressure, which sits in between the regimes
in which the disc is dominated by gas pressure (lower accretion rates) and in which
advection starts to play a major role (the highest accretion rates), but the quantitative
range of accretion rates over which \rmag\ remains constant depends on the
magnetic field of the neutron star.}, and so any propeller transitions that occur in this
regime may have to invoke changes in \rco. Having propeller transitions can be
driven by changes in \rco\ would remove the expectation that such transitions must
be seen at lower luminosities. 

Although the spin-up rates in ULX pulsars can be large (\eg\ \citealt{Bachetti14nat,
Carpano18}), potentially resulting in appreciable changes in \rco\ on observable
timescales, even in these systems the spin periods ($P$) do not change by large
factors over the $\sim$decade timescales they have been observed
(\citealt{Bachetti20, Bachetti22, Fuerst21})\footnote{NGC\,300 ULX1 is a notable
exception, which does change its spin period significantly on these timescales
(\citealt{Vasilopoulos19}).}. Furthermore, \rco\ has a moderately shallow dependence
on $P$ (it should scale as \rco\ $\propto P^{2/3}$). Invoking a propeller transition
driven by changes in $P$ would place a strong requirement that the source
was already extremely close to spin equilibrium (\ie \rco\ $\simeq$ \rmag). It is also
unclear whether accretion at a relatively steady rate could genuinely spin-up the
neutron star to the point where it enters the propeller regime. As the neutron star
approaches spin equilibrium the spin-up rate should taper off (as the disc can impart
less and less angular momentum), meaning that \rco\ may only be able to asymptote
towards \rmag, but not cross it, if only $P$ is changing.  It may be easier, instead, to
invoke changes in the orbital frequencies within the disc, e.g. due to changing
viscosity, as a mechanism for inducing a propeller transition by changing \rco.
However, the similar soft X-ray fluxes ($\lesssim$2\,keV) seen in the soft state
epochs to the other high-flux epochs would suggest there are not large changes in
accretion rate through the disc, so this would likely also require \rco\ $\simeq$ \rmag.
Given all of this, it is not clear whether the propeller regime really offers a plausible
scenario for the soft state observations in \hoix; this might only be possible under
very specific circumstances.

\label{lastpage}

\end{document}